\documentclass[pdftex,twocolumn,epjc3]{svjour3}          % twocolumn

\RequirePackage[T1]{fontenc}

\smartqed  % flush right qed marks, e.g. at end of proof

\RequirePackage{graphicx}
\RequirePackage{subcaption}
\RequirePackage{float}
\RequirePackage{mathptmx}      % use Times fonts if available on your TeX system
\RequirePackage{flushend}
\RequirePackage[numbers,sort&compress]{natbib}
\RequirePackage[colorlinks,citecolor=blue,urlcolor=blue,linkcolor=blue]{hyperref}
\RequirePackage{newtxmath}
\RequirePackage{tensor}

\journalname{Eur. Phys. J. C}

\begin{document}

\title{Peculiar velocity fields from analytic solutions of General Relativity}

\subtitle{}

\author{Roberto A. Sussman\thanksref{e1,addr1}
        \and
        Sebasti\'an N\'ajera\thanksref{e2,addr2} \and F A Piza\~na\thanksref{addr1} \and Juan Carlos Hidalgo\thanksref{addr2} %etc.
}

%\thankstext[$\star$]{t1}{Thanks to the title}
\thankstext{e1}{e-mail: sussman@nucleares.unam.mx}
\thankstext{e2}{e-mail: sebastian.najera@icf.unam.mx}

\institute{Instituto de Ciencias Nucleares, Universidad Nacional Aut\'onoma de M\'exico
(ICN--UNAM),\\ A. P. 70 –- 543, 04510 M\'exico D. F., M\'exico.\label{addr1}
          \and
          Instituto de Ciencias F\'isicas, Universidad Nacional Aut\'onoma de M\'exico, 62210, Cuernavaca, Morelos.\label{addr2}
}

\date{Received: date / Accepted: date}

\maketitle

\begin{abstract}
Peculiar velocities are analyzed through cosmological perturbations in the Newtonian longitudinal gauge characterized by irrotational shear-free congruences in an Eulerian frame. We show that non-trivial peculiar velocity fields can be generated through Lorentzian boosts in the non-relativistic limit, where the Eulerian frame is obtained from analytic solutions of Einstein's equations sourced by an irrotational shear-free fluid with nonzero energy flux. This approach provides a physically viable interpretation of these analytic solutions, which (in general) admit no isometries, thus allowing, in principle, for modeling time and space varying 3-dimensional fields of peculiar velocities that can be contrasted with observational data on our local cosmography. As a ``proof of concept'' we examine the peculiar velocities of varying dark matter and dark energy perfect fluids with respect to the CMB frame using a simple, spherically symmetric particular solution. The resulting peculiar velocities are qualitatively compatible with observational data on the CMB dipole.    
\end{abstract}

\section{Introduction}\label{intro}

%\section{Introduction}

Peculiar velocities are important in cosmological research, as they convey the  relative motion of local self-gravitating systems with respect to the Hubble flow and also relative internal motions within large scale structures \cite{turner2024cosmology,carrick2015cosmological}.  Given their local and non-relativistic nature, the simplest approach to predict and compute these  velocities is to treat them as relative velocities from a purely Doppler redshift at very low redshift, $z\ll 1$, from a Galilean subtraction of local motion from the Hubble ``bulk'' flow \cite{davis2014deriving,boruah2020cosmic}. Another approach is the reconstruction of a 3-dimensional velocity field from observations \cite{boruah2021peculiar,wang2024peculiar,dupuy2023dynamic}.  

Peculiar velocities are closely related to density perturbations and redshift space distortion \cite{peebles2020principles,ellis2012relativistic}.  While a purely Newtonian treatment of peculiar velocities is appropriate at very low redshifts, relativistic corrections are needed as observations probe increasingly larger scales, comparable to the Hubble horizon. Within a general relativistic  framework they can  be understood as non-relativistic relative velocities between sources along different frames \cite{tsagas2025large}: matter-energy sources (baryons, Cold Dark Matter (CDM), dark energy) that are not comoving with a frame that can be identified with the Cosmic Microwave Background (CMB) frame or with a given FLRW model, such as the $\Lambda$CDM FLRW background. This conjecture can be tested \cite{kraljic2016frames,kashlinsky2022probing} and is also part of the controversy surrounding the Cosmological Principle  \cite{tsagas2025large,aluri2023observable,rameez2021there,secrest2022challenge}.

A relativistic framework for modeling peculiar velocities at large scales  is furnished by gauge invariant linear cosmological perturbations on an FLRW background. In the gauge invariant metric formalism  peculiar velocities are more conveniently analyzed in the Newtonian longitudinal gauge, characterized by irrotarional shear-free congruences \cite{ma1995cosmological,burgazli2020effect,cembranos2019non}. An equivalent  relativistic perturbative approach is the covariant formalism of Ellis and Bruni, which considers  an Eulerian frame defined by an irrotational  shear-free 4-velocity field equivalent to the congruences of the Newtonian gauge, with  peculiar velocities obtained from a Lorentzian boost from sources comoving in a Lagrangian frame \cite{maartens1998covariant,ellis2001general}.  The relativistic study of peculiar velocities has so far been conducted almost exclusively within a perturbative approach \cite{tsagas2025large,aluri2023observable,burgazli2020effect,cembranos2019non,wagenveld2023cosmic,miliou2024peculiar}. 

A non-perturbative analytic and numerical relativistic approach to non-comoving 4-velocity frames ({\it i.e.} ``tilted'') has been undertaken with Bianchi models  \cite{coley2005dynamical,shogin2015dynamics}, spherically symmetric \cite{bolejko2025,delgado2019,pizana2023} and Szekeres \cite{najera2021non} models. However, since relativistic perturbation formalisms assume Eulerian irrotational shear-free  congruences or frames, a more  appropriate non-perturbative relativistic approach, consistent with cosmological perturbations could be based on analytic or numerical solutions of Einstein's equations of spacetimes whose sources are  irrotational shear-free fluids. 

Exact solutions whose source is an irrotational shear-free perfect fluid are among the oldest  known  solutions of Einstein's equations \cite{sussman1988,krasisnki1989,collins1986shear, carminati1987shear, ellis2011shear, zaeem2016shear}.  We can rule out these perfect fluid exact solutions as non-perturbative analogues of Eulerian frames to analyze peculiar velocities, since a perfect fluid on top of zero shear places overly restrictive constraints to allow for the description of fluids with different 4-velocities.  However,  exact  solutions are also known with irrotational shear-free imperfect fluids compatible with nonzero energy flux (off-diagonal terms in the energy momentum tensor). Although the most general known solution of this type admits no isometries \cite{sussman1993new}, practically all existing literature on these solutions regard the energy flux as thermal heat conduction in  simplified toy models of radiative collapse in a Vaidya background    \cite{de1985collapse,santos1985non, bonnor1989radiating,grammenos1995radiating, majozi2025complexity}.   

However, energy flux as heat conduction  is only valid in a hydrodynamical regime relying on short range interactions in collisional fluids, which is a completely inappropriate theoretical framework for the long range largely non-collisional interactions of gravity dominated systems. Therefore, following the covariant approach in previous literature  \cite{maartens1998covariant,ellis2001general,tsagas2010large}, we consider energy flux only in connection with a non-perturbative approach to non-relativistic peculiar velocities  arising from Lorentzian boosts between non-comoving fluids.  

With this aim, we begin by summarizing the  most general known exact solution for an irrotational shear-free fluid with nonzero energy flux  derived in 1993 \cite{sussman1993new}, showing that its ``imperfect fluid'' energy momentum tensor can be exactly rewritten as a mixture of two non-comoving perfect fluids, with the energy flux given in terms of peculiar velocities that emerge from the non-relativistic limit of a Lorentzian boost between the two different 4-velocities.  

Since the general solution admits no isometries and allows for (up to) 9 free functions, it has the potential (in principle) to model 3-dimensional vector fields of peculiar velocities that can be  matched or compared with observational data \cite{aluri2023observable,wagenveld2023cosmic,jarrett20002mass}, specially peculiar velocity surveys \cite{boruah2020cosmic,sorrenti2023dipole} and the Cosmicflows-4 \cite{hoffman2024large}. However, these are comprehensive and laborious tasks that require separate dedicated articles that will be undertaken in the future. Therefore, as a ``proof of concept'', we show how with a simple particular solution it is possible to achieve a reasonable (but simplified) modeling of inhomogeneous and time varying peculiar velocities of matter (baryons and CDM) and varying dark energy with respect to the CMB frame, with correspondence with redshift and distance estimations normally used in perturbative studies.    

The paper is organized as follows. In Sec.~\ref{sec:gensol} we introduce and describe the most general known  solution of Einstein's equations for an irrotational shear-free fluid. In Sec.~\ref{sec:physical} we provide a physical interpretation of the energy flux of this solution in terms of non-relativistic peculiar velocities arising from a Lorentzian boost between non-comoving perfect fluids. We present in Sec.~\ref{sec:simpmod}  a simplified particular solution, showing in Sec.~\ref{sec:near} that it leads to nearly homogeneous models whose dynamics can be  described by  series expansions around a closed FLRW background. Sec.~\ref{sec:flumix} discusses the interpretations of these models as mixtures of non comoving fluids, presenting two physically motivated scenarios involving inhomogeneous matter and dark energy. In Sec.~\ref{sec:numres} we discuss the physical properties of the fluid mixtures and present numerical results obtained for peculiar velocities and their connection with local redshifts. Finally, Sec.~\ref{sec:conc} summarizes our findings and discusses prospects for extending the present work to more general solutions and observational applications. We provide two appendices, \ref{sec:coords} discusses regularity conditions, coordinate transformations and geometrical properties of the spherically symmetric models, while \ref{sec:series} assesses the accuracy of the series expansion by comparison with exact expressions. 

\section{The most general known solution}\label{sec:gensol}

The most general known exact solution whose source is an irrotational shear-free fluid was derived in 1993 \cite{sussman1993new}. It is described by the following metric and energy momentum tensor in a comoving frame
\footnote{This  is not the most general solution, since the rest frames (3-dimensional hypersurfaces orthogonal to $u^a$) are conformally flat. In the most general solution anisotropic pressure is not zero and the rest frames are conformal to an arbitrary 3 dimensional metric  $g_{ij} = L^{-2} h_{ij}$ where the $h_{ij}$ do not depend on time. } 
\begin{eqnarray}ds^2&=&\frac{-N^2 dt^2 +\delta_{ij} dx^i dx^j}{L^2},\label{metric1}\\
T^{ab}&=&\rho u^au^b +ph^{ab}+2q^{(a}u^{b)},\qquad u^a=\frac{L}{N} \delta^a_t.\label{Tab}\end{eqnarray}
where $N=N(t,x^i),\,\,L=L(t,x^i)$. The nonzero kinematic parameters are the  expansion ($\Theta = \nabla_a u^a$) and 4-acceleration ($A_a=\dot u_a = u^b\nabla_b u_{a}$)
\begin{equation}\Theta= \frac{3L_{,t}}{N},\qquad A_i=\dot u_i=\left(\ln \frac{N}{L}\right)_{,i},\quad \dot A_0=u_0=0. \label{ThAcc} \end{equation}
The first order 1+3 covariant formulation of Einstein's equations leads to four evolution equations 
\begin{eqnarray}
\dot \rho&=&-(\rho+p)\Theta -\tilde\nabla_a q^a-2A_aq^a,\label{ev1}\\
\dot \Theta &=& -\frac13\Theta^2-4\pi(\rho+3p)+\tilde\nabla_a A^a+A_a A^a,\label{ev2}\\
\dot q_{\langle a\rangle}&=&-\frac43\Theta q_a -(\rho+p)A_a -\tilde \nabla_a p,\label{ev3}\\
\dot E_{\langle ab\rangle}&=&-\Theta E_{ab}+4\pi\tilde \nabla_{\langle a}q_{b\rangle}-8\pi q_{\langle a}A_{b\rangle},\label{ev4}
\end{eqnarray}
and the following 4 constraints
\begin{eqnarray}
\tilde\nabla_b E^b\,_a&=&\frac{8\pi}{3}\left(\tilde \nabla_a\rho-\Theta q_a\right),\label{con1}\\
E_{ab}&=&\tilde\nabla_{\langle a}A u_{b\rangle}+A_{\langle a}A_{b\rangle},\label{con2}\\
8\pi q_a &=& \frac23 \tilde \nabla _a\Theta,\label{con3}\\
\frac{8\pi}{3}\rho &=& \frac{\Theta^2}{9}+\frac16{}^3 R,\label{con4}
\end{eqnarray}
where the angle brackets ${}_{\langle ..\rangle}$ denote spacelike projection, symmetrization and removal of the trace, and $\tilde\nabla_a=\tensor{h}{_a ^b}\nabla_b$ is the covariant derivative in the rest frames. The three constraints \eqref{con2},  \eqref{con3} and  \eqref{con4} respectively furnish definitions of the electric Weyl tensor $E_{ab}$, the energy flux vector $q_a$ and the density $\rho$, which can be readily computed from \eqref{ThAcc} and by computing ${}^3 R$ for the conformally flat 3=metric:
\begin{equation}\frac16 {}^3 R =\delta^{ij}\left( \frac23 L_{, ij}\,L- L_{,i}L_{,j}\right).\end{equation}
These are not the only constraints. The fact that three components of $T_{ab}$ vanish ($T_{xy}=T_{yz}=T_{zx}=0$) plus the vanishing of the anisotropic pressure ($\Pi^{ab}=0$), leads to the following six extra constraints
\begin{equation} G_{ij}=0,\,\,\, i\ne j,\qquad G_{ii}-G_{jj}=0,\,\,i,j=1,2,3\,\,\hbox{no sum}.\label{sixeqs}\end{equation}
As proven in \cite{sussman1993new}, the most general known analytic solution of the system  \eqref{sixeqs} follows by the metric functions $N$ and $L$ taking the form
\begin{equation} N= N(\Phi),\quad L=L(\Phi), \label{LNPhi}\end{equation}
where  $\Phi=\Phi(t,x^i)$ satisfies
\begin{eqnarray}   \Phi_{,ij}&=& 0,\quad i\ne j, \quad \Phi_{,ii}=0,\quad i=x,y,z,\nonumber\\
\Phi&=&\alpha(t) +\beta_i(t) x^i+\gamma(t)\delta_{ij}x^i x^j,\label{Phi}\end{eqnarray}
and $L,\,N$ satisfy the following coupled linear differential equations:
\begin{equation} \frac{d^2 N}{d\Phi^2} = J(\Phi) N, \qquad 2\frac{d^2 L}{d\Phi^2} = J(\Phi) L,\label{2linear}\end{equation}
where $J(\Phi)$ is an arbitrary function related to the conformal invariant $\Psi_2$  \cite{sussman1993new}
\begin{equation} J=\frac{6\ell^2 \Psi_2 }{\Phi^2 L^2}\qquad \Rightarrow\quad \Psi_2 =\frac{\ell^2 J\,\Phi^2 L^2}{6}.\label{Psi2} \end{equation}
where $\ell$ is an inverse length scale. The function $J(\Phi)$ also has an invariant characterization in terms of the Weyl tensor
\begin{equation}\tensor{C}{_a_b^c^d}=4\left(\tensor{h}{_[_a^[^c} u_{[a}u^{[c} \right)\tensor{E}{_b_]^d^]},\quad 
E_{ab}=\frac12 J(\Phi)\nabla_{\langle a}\nabla_{b\rangle}\Phi,\label{Weyl} \end{equation}
which also relates it to the 4-acceleration through \eqref{con2}. The choice of the generatrix function $J(\Phi)$ determines specific exact solutions.  Once $N$ and $L$ satisfy \eqref{2linear} the variables $\rho,\,p$ and $q_a$ in the field equations are
\begin{eqnarray}
  \frac{8\pi}{3}\rho &=&\frac{L'^2}{N^2}\Phi_{,t}^2+\left(\frac13 J-\frac{L'^2}{L^2}\right)\,L^2 \delta^{ij} \Phi_{,i}\Phi_{,j},\label{eqrho_}\\
  8\pi p &=& \frac{2L L'}{N^2}\,\Phi_{,tt}+\frac{L^2}{N^2}\left[J-\frac{L'}{L}\left(\frac{3L'}{L}+\frac{2N'}{N}\right)\right]\,\Phi_{,t}^2\nonumber\\
  &&-\left(\frac{2N'}{N} -\frac{3L'}{L}\right)\,L L'\delta^{ij}\Phi_{,i}\Phi_{,j},\label{eqp_}\\
  8\pi q^a   &=& 8\pi Q^i\delta_i^a,\nonumber\\
  8\pi Q^i&=&-\frac{L^2}{N^2}\left[\left(J\,N\,L-2\,N'\,L'\right)\Phi_{,i}\Phi_{,t}+2\,N\,L'\,\Phi_{,it}\right],\label{eqQ_}
\end{eqnarray}
where the primes denote $d/d\Phi$.   

The solutions of \eqref{2linear} produce four extra functions of $t$ as integration constants, plus the four arbitrary functions in $\Phi$ adds to ten free functions of $t$. Even before solving \eqref{2linear}, the linear system allows one to eliminate second order derivatives of $N$ and $L$ and once  \eqref{2linear} is solved for a specific choice of $J(\Phi)$ the parameters $\Theta$ and $\dot u_a$ in \eqref{ThAcc} can be computed and then $E_{ab},\,q_a$ and $\rho$ follow from  \eqref{con2}-\eqref{con4} and \eqref{Weyl}.

\section{Physical interpretation: peculiar velocities}\label{sec:physical}

As long as the energy flux vector $q^a$ in \eqref{Tab} and \eqref{eqQ_} is interpreted as dissipative heat conduction, the physical interpretation of the shear-free solutions \eqref{metric1}-\eqref{Tab} in a comoving frame is elusive, as there is a high likelihood that the state variables \eqref{LNPhi}-\eqref{eqQ_} will be incompatible with the appropriate thermal equations of state and constitutive equations. 

Heat conduction is not an appropriate interpretation of $q^a$ in a cosmological context dominated by long range gravitational interaction. Instead, in consistency with the literature on the relativistic cosmological approach to non-relativistic peculiar velocities,  we propose a physical interpretation of the solutions by considering $q^a$ in connection with a peculiar velocity field that results from relating two distinct 4-velocities $\hat u^a$ and $u^a$ related by a Lorenzian boost: .   
\begin{equation}\hat u^a =\gamma(u^a + v^a),\qquad \gamma = \frac{1}{\sqrt{1-v_av^a}},\quad v_a u^a =0. \label{boost}\end{equation}
Consider a perfect fluid in the frame of $\hat u^a$ (Lagrangian frame)
\begin{equation} \hat T^{ab} = (\hat \rho+\hat p) \hat u^a \hat u^b + \hat p g^{ab},\label{hatPF}\end{equation} 
that is not comoving with the frame of $u^a$ (Eulerian frame). Applying  \eqref{boost} to \eqref{hatPF}, the energy momentum tensor in the frame of  $u^a$ is (see the transformation for arbitrary $T^{ab}$ in \cite{tsagas2025large,maartens1998covariant})
\begin{eqnarray} 
T^{ab} &=& (\rho+p) u^a u^b + p g^{ab} + 2 q^{(a}u^{b)}+ \Pi^{ab}\label{nohat},\\
\rho &=& \hat \rho+\gamma^2 v_av^a (\hat \rho + \hat p),\quad p= \hat p+\frac13\gamma^2 v_av^a (\hat \rho + \hat p),\label{hatrhop}\\
q^a &=& \gamma^2 (\hat \rho+\hat p) v^a,\quad \Pi^{ab}=\gamma^2 (\hat \rho+\hat p) v^{\langle a}v^{b\rangle},\label{hatQPi}
\end{eqnarray}
where $v^{\langle a}v^{b\rangle}$ dennotes the spatial trace-free tensor product. For non-relativistic peculiar velocities we have  $v_av^a\ll 1,\,v^{\langle a}v^{b\rangle} \ll 1$ and thus, up to $\mathcal{O}(v/c)^2$, we have
\begin{equation}\gamma\approx 1, \quad \hat \rho \approx \rho,\quad  \hat p \approx p,\quad q^a \approx (\hat \rho+\hat p) v^a\approx (\rho+p) v^a,\label{linappr}\end{equation}
leading to the following total energy momentum tensor in the Eulerian frame    
\begin{equation} T^{ab}=(\rho +p)u^au^b+p g^{ab} + 2(\rho+p) v^{(a} u^{b)},\label{nohatTab}\end{equation}
which  provides a physically appealing interpretation to the energy flux $q^a$ in \eqref{Tab} that is (in principle) consistent with considering these solutions as cosmological models. 

The conservation equations for energy momentum tensor \eqref{hatPF} are those of a perfect fluid in the frame defined by the 4-velocity $\hat u^a$ 
\begin{equation} \dot{\hat\rho}+(\hat \rho+\hat p)\hat\Theta=0,\quad \tilde \nabla_a\hat p+(\hat\rho+\hat p)\hat A_a=0,\label{hatcons}\end{equation}
where \eqref{linappr} holds in the non-relativistic approximation $v^a/c\ll 1$, but the kinematic parameters \eqref{ThAcc} associated to the 4-velocity $\hat u^a$ in \eqref{hatPF} are also modified by the boost transformation \eqref{boost}. In the non-relativistic limit they take the form (the general form is given in \cite{tsagas2025large,maartens1998covariant})
\begin{equation} \hat \Theta =\Theta+\tilde \nabla_a v^a, \quad \hat A_a = A_a +\dot v_a+\Theta v_a,\quad \hat \sigma_{ab}=\tilde\nabla_{\langle a}v_{b\rangle}.\end{equation}
where $\dot v^a =u^b\nabla_b v^a$. Notice that $\hat u^a$ is not a shear-free 4-velocity even if $u^a$ is shear-free. 

For the solutions \eqref{metric1}-\eqref{Tab}, we have, at first order, $\hat \rho$ and $\hat p$ given by \eqref{eqrho_} and \eqref{eqp_}, while the 3-dimensional field of non-relativistic peculiar velocities is
\begin{eqnarray}   v^a &=& v^i\delta_i^a,\nonumber\\
 v^i &=& \frac{Q^i}{\rho+p}\nonumber\\
 &=& -\frac{N\,L\,\left[2NL' \Phi_{,it}+\Phi_{,t}\Phi_{,i}(NJL-2L' N')\right]}{2NL'/L\Phi_{,tt}+(N^2\delta^{ij}\Phi_{,i}\Phi_{,j}+\Phi_{,t}^2)(NJL-2L' N')}.\nonumber\\
 \label{pecvel}\end{eqnarray}
Evidently, the interpretation of $q^a$ as an energy momentum  flux associated with non-relativistic  peculiar velocities in \eqref{pecvel} obtained from the general metric \eqref{metric1} is potentially useful for modeling the evolution of 3-dimensional velocity fields that can be fitted to observations in peculiar velocity surveys. However, the actual derivation of these models still requires solving the system \eqref{2linear} and looking at the relation between  $\rho$ in \eqref{eqrho_} and $p$  in \eqref{eqp_} for an appropriate assumption of fluid mixtures, a task that will be addressed in future work. In order to illustrate the viability of this process, we consider a simplified spherically symmetric example. 

\section{Simplified models: conformally flat non-accelerating fluids}\label{sec:simpmod}

We need  to solve  the system \eqref{2linear} in order to provide a simple illustrative example to be able to show that the solutions presented in sections \ref{sec:gensol} and \ref{sec:physical} can lead to physically motivated cosmological models. For this purpose, we consider the simplest possible solutions of \eqref{2linear} characterized by
\begin{itemize}
\item Conformal flatness follows from setting $J(\Phi)=0$, leading from \eqref{Weyl} to a vanishing Weyl tensor (the Petrov type O particular case). These are the simplest solutions of \eqref{2linear}:
\begin{equation}   \frac{d^2 N}{d\Phi^2} =  \frac{d^2 L}{d\Phi^2} = 0\; \Rightarrow\quad L(\Phi)=\lambda_1 +\lambda_2 \Phi,\; N(\Phi)=\nu_1 +\nu_2 \Phi,\label{2linearCF}\end{equation}
where $\lambda_1,\,\lambda_2,\,\nu_1,\,\nu_2$ depend on time and $\Phi$ is given by \eqref{Phi}.
\item Non-accelerating fluid. The 4-acceleration vanishes if (from  \eqref{ThAcc}) we choose,
\begin{equation}  \lambda_1=\nu_1,\,\,\textrm{and}\,\,\lambda_2=\nu_2\; \Rightarrow\; N=L \quad \Rightarrow\; \dot u_a =\tilde \nabla_a\left(\ln \frac{N}{L}\right)=0,\label{noacc}\end{equation}
leading to the metric \eqref{metric1} with $g_{tt}=-1$ and $L$ given by \eqref{2linearCF}. 
\item Spherically symmetric sub-cases of the solutions derived in section \ref{sec:gensol} follow by setting the three free functions $\beta_i(t)=0$ in \eqref{Phi}, so that $\Phi=\alpha(t) +\beta(t) r^2,$ with $r^2=x^2+y^2+z^2$.     
\end{itemize}
However, the particular cases of \eqref{metric1}-\eqref{Tab}  with conformal flatness, zero acceleration and spherical symmetry are not the most general solutions satisfying these properties (derived by Coley and McMannus \cite{coley1994}). Since in these particular cases of \eqref{metric1}-\eqref{Tab} the rest frames are conformally flat, we can consider more general spherically symmetric solutions in which the rest frames are conformal to 3-dimensional hypersurfaces with constant curvature (which includes conformal flatness as the case of zero spatial curvature). 

The spherically symmetric models satisfying \eqref{2linearCF}-\eqref{noacc} that provide a more  illustrative example of the construction and study of peculiar velocities, as described in Section~\ref{sec:physical}, are characterized by the following metric (the justification and  coordinate transformations leading to this metric are discussed in Appendix \ref{sec:coords}):
\begin{eqnarray}    ds^2 &=& -dt^2 +\frac{a^2(t)\left[d r^2+f^2(r)(d\theta^2+\sin^2\theta d\phi^2)\right]}{L^2},\label{metricCF2}\\
L&=&1+ b(t)F(r),\nonumber\\
f(r)&=&\sin r,\qquad F(r)=1-\cos r=2\sin^2 r/2,\label{Fchi2}\end{eqnarray}
where $b(t)$ is a dimensionless free function.  Einstein's equations $G_{ab}=8\pi T_{ab}$ for the metric \eqref{metricCF2} and the  energy momentum tensor \eqref{Tab} are: 
\begin{eqnarray}
  \frac{8\pi\rho}{3} &=& \frac{\dot a^2}{a^2}+\frac{k_0}{a^2}+\frac{\dot b^2}{b^2}-\frac{2\dot a\dot b}{ab}+\frac{2\epsilon_0b}{a^2}\nonumber\\
  &&-\frac{\dot b\left[(2a\dot b-2b\dot a)(1+\epsilon_0bF)-a\dot b\right]}{ab^2(1+\epsilon_0bF)^2},\label{eqrhoCF1}\\
  8\pi p&=&-\frac{\dot a^2}{a^2}-\frac{2\ddot a}{a}-\frac{k_0}{a^2}+\frac{\ddot b}{b}+\frac{\dot b^2}{b^2}+\frac{6\dot a\dot b}{ab}-\frac{2\epsilon_0b}{a^2}\nonumber\\
&=&-\frac{\dot b\left[((10\dot b^2-2b\ddot b)a-6\dot a\dot b)(1+\epsilon_0bF)-5\dot b^2\right]}{ab^2(1+\epsilon_0 bF)^2},\label{eqpCF1}\\
  8\pi q^a &=& 8\pi Q \delta^a_r,\qquad 8\pi Q =- \frac{2\dot b\,f}{a^2},\label{eqQCF1}
\end{eqnarray}
where the dot denotes coordinate time derivative: $\dot b=b_{,t}$ (which coincides with proper time derivative) and the  constant $k_0>0$ is an inverse squared length related to the spatial curvature by
\begin{equation}\frac16 {}^3R = \frac{k_0\,L^2}{a^2}+\frac{b(2f' L-b)}{a^2},\label{spatcurv}\end{equation}
which reduces to the spatial curvature of ``closed'' FLRW models $k_0/a^2$ in the FLRW limit $b(t)=0,\,\, L=1$. As we discuss in \ref{sec:coords}, these models comply with  regularity conditions and provide an appropriate generalization of ``closed'' FLRW models.  

\section{Nearly homogeneous models}\label{sec:near}

In models described by \eqref{metricCF2}-\eqref{Fchi2} we have $0\leq F(r)\leq 2$, while  choosing a smooth function $b(t)$ such that $0\leq b(t)\leq \epsilon_0$ holds for all $t$, for a sufficiently small $\epsilon_0$ (even if negative), then:  $0\leq \epsilon_0\,b\,F\leq 2\epsilon_0$ remains bounded over the full coordinate range $(t,\,r)$. Consequently, we can redefine $b$ as $\epsilon_0\,b$ and consider  series expansions of the metric and all variables (including \eqref{eqrhoCF1}-\eqref{eqQCF1}) around a sufficiently small $\epsilon_0$. In particular, the metric \eqref{metricCF2} becomes at first order in $\epsilon_0$ 
\begin{eqnarray}ds^2=-dt^2+a^2(t)\left(1-2\,\epsilon_0\,b\, F\,\right)\left[dr^2+f^2\left(d\theta^2+\sin^2\theta d\phi^2\right)\right],\nonumber\\
\label{metricCF3}\end{eqnarray}
denoting a metric of inhomogeneous models that are close (up to order $\epsilon_0$) to the metric of a reference FLRW background model with positive spatial curvature (the limit $\epsilon_0=0$).  

Since $q^a$ in \eqref{eqQCF1} is already of order $\epsilon_0$ (directly proportional to $\dot b$), we only expand around $\epsilon_0$ the density and pressure in \eqref{eqrhoCF1}-\eqref{eqQCF1} and the Hubble expansion scalar $\Theta=\nabla_a u^a$ (the only nonzero kinematic parameter)
\begin{eqnarray}
\frac{8\pi}{3}\rho &=& \frac{\dot a^2}{a^2}+\frac{k_0}{a^2}+\frac{2[b-(a\dot a \dot b)\,F]}{a^2}\,\epsilon_0+O(\epsilon_0^2),\label{eqrhoCF2}\\
8\pi p &=& -\frac{\dot a^2}{a^2}-\frac{2\ddot a}{a}-\frac{k_0}{a^2}+\frac{2[(\ddot ba+3\dot a \dot b)\,a\,F-b]}{a^2}\,\epsilon_0\nonumber\\
&&+O(\epsilon_0^2),\label{eqpCF2}\\
8\pi Q &=&- \frac{2\epsilon_0\dot b\,f}{a^2},\label{eqQCF2}\\
\frac{\Theta}{3}&=&\frac{\dot a}{a}-\frac{\epsilon_0\,\dot b\,F}{1+\epsilon_0\,b\,F}\approx \frac{\dot a}{a}-\epsilon_0\,\dot b\,F.\label{eqThCF2}
\end{eqnarray}
Besides simplifying the expressions, the series expansions  \eqref{eqrhoCF2}-\eqref{eqThCF2} provide an excellent and very accurate approximation to the exact expressions for $\epsilon_0 < 0.1$ (we prove this in \ref{sec:series}). Equations \eqref{eqrhoCF2}-\eqref{eqThCF2} also  highlight the role of closed  FLRW models as reference spacetimes, since only $\rho,\,p$ and $\Theta$ (common to FLRW) have zeroth order terms involving $a(t)$, while  the second time dependent function $b(t)$ only appears at order $\epsilon_0$ and thus controls the time evolution of the corrections to the background reference.  

Notice that even at the symmetry centers $r=0,\,\pi$ the zeroth order FLRW equations in \eqref{eqrhoCF2}-\eqref{eqpCF2} do not fully determine the time evolution, since the terms of order $\epsilon_0$ in \eqref{eqrhoCF2} and \eqref{eqpCF2} contain the purely time dependent function $2\epsilon_0 b/a^2$.    
   
\section{Fluid mixtures}\label{sec:flumix}
 
The energy momentum tensors \eqref{hatPF} and \eqref{nohat} related by the boost \eqref{boost} under the non-relativistic approximation \eqref{linappr} can each represent fluid mixtures in different 4-velocity frames that are compatible with the field equations \eqref{eqrhoCF2}-\eqref{eqQCF2}. We can write the density and pressure in \eqref{eqrhoCF2} and \eqref{eqpCF2} as
\begin{equation}
 \rho= \rho_{(0)}(t)+\epsilon_0\,\delta_\rho(t,r),\qquad p= p_{(0)}(t)+\epsilon_0\,\delta_p(t,r)
\end{equation}
where $\rho_{(0)},\,\,p_{(0)}$  are the zeroth order FLRW equations and $\delta_\rho,\,\,\delta_p$  the order $\epsilon_0$ terms in \eqref{eqrhoCF2} and \eqref{eqpCF2} (including time dependent terms with $b(t)$).  We will consider the relative motion with respect to the CMB reference frame, as this photon gas is a primordial relic source of the radiation plasma emerging from the early Universe, well before structure formation. 

To determine the homogeneous fluids (CMB and the zeroth order term of matter) we consider the FLRW dynamical equations at zeroth order  in \eqref{eqrhoCF2} and \eqref{eqpCF2}     
\begin{equation} 
  \frac{\dot a^2}{H_0^2}=\frac{\Omega_0^{(r)}}{a^2}+\frac{\Omega_0^{(m)}}{a}-\Omega_0^K+\Omega_0^{\tiny{\Lambda}}\,a^2,\; \frac{\ddot a}{H_0^2}=\frac{2\Omega_0^{\tiny{\Lambda}}\,a^4-\Omega_0^{\tiny{\textrm{m}}}\,a-2\Omega_0^{(r)}}{2a^3},\label{zeroO}\end{equation}
where we have now identified $k_0 = H_0^2\Omega_0^K = H_0^2(\Omega_0^{(r)}+\Omega_0^{(m)}+ \Omega_0^{\tiny{\Lambda}}-1)$. To determine the terms of order $\epsilon_0$ we need to specify the function $b(t)$, which ideally should follow from assumptions on the fundamental properties of a dark matter or dark energy models, a task whose consideration is outside the purpose of this paper. Therefore, we consider the following dimensionless bounded and decaying function
 \begin{equation} b(t) = \exp (-a^2(t)).\label{bfun}\end{equation}
In what follows we assume that $\hat u^a\approx u^a+v^a$ and $\hat \rho^{(m)}\approx \rho^{(m)},\,\,\hat p^{(m)}\approx p^{(m)}$ hold under the non-relativistic approximation \eqref{linappr}. We consider the following two fluid scenarios. 
%
%\begin{itemize}
%\item 

\subsection{Scenario 1: inhomogeneous non-comoving matter}

The Lagrangian frame $\hat T_{\tiny{\hbox{}L}}^{ab}$ corresponds to matter: baryons and cold dark matter (CDM) as an inhomogeneous perfect fluid, while the Eulerian frame $T_E^{ab}$: the CMB radiation as a homogeneous perfect fluid (a photon gas) with dark energy as cosmological constant
\begin{eqnarray}
 \hat T_{\tiny{\hbox{L}}}^{ab} &=& \hat\rho^{(m)} \hat u^a\hat u^b +\hat p^{(m)} \hat h^{ab},\nonumber\\
 \hat\rho^{(m)}&=&\hat\rho_{(0)}^{(m)}+\epsilon_0\,\delta^{(\hat\rho)},\; \hat p^{(m)}=\epsilon_0\,\delta^{(\hat p)},\\
 T_{\tiny{\hbox{E}}}^{ab}&=&(\rho^{(r)} +\Lambda) u^a u^b +(p^{(r)}-\Lambda) h^{ab},\; p^{(r)}=\frac13\rho^{(r)}, \end{eqnarray}
with the total energy momentum $T^{ab}=T_{\tiny{\hbox{L}}}^{ab}+T_{\tiny{\hbox{E}}}^{ab}$ in the Eulerian frame given by
\begin{eqnarray} T^{ab}&=&(\rho^{(r)}+\rho^{(m)}+\Lambda)u^au^b+(p^{(m)}-\Lambda)h^{ab}+ 2q^{(a}u^{b)},\nonumber\\
\label{TabE}
\end{eqnarray}
where
\begin{eqnarray} q^a &=& [\rho^{(m)}+p^{(m)}]v^a=[\rho_{(0)}^{(m)}+\epsilon_0(\delta^{(\rho)}+\delta^{p)})]\,v^a.\label{TabE2}\end{eqnarray} 
%
%\end{itemize}
%
 The constituent fluids are 
 \begin{itemize}
 \item Lagrangian frame $\hat T_{\tiny{\hbox{L}}}^{ab}$ (baryons and CDM):
 \begin{eqnarray} \frac{8\pi\,\hat\rho}{3 H_0^2}&=&\frac{\Omega_0^{(m)}}{a^3}+\epsilon_0\Delta^{(\rho)},\qquad \frac{8\pi\,\hat p}{3 H_0^2}=\epsilon_0\Delta^{(p)},\label{TLrhop}\\
  \Delta^{(\rho)} &=&\frac{8\pi}{3H_0^2} \delta^{(\rho)}=\frac{2\,\left[1+2 F{\cal B}_1(a)\right]\,e^{-a^2}}{a^2},\label{TLrho}\\
   \Delta^{(p)} &=&\frac{8\pi}{3H_0^2} \delta^{(p)}\nonumber\\
   &=&\frac{2\,
 \left[1+(4(2+a^2){\cal B}_1(a)+{\cal B}_2(a))F\right]\,e^{-a^2}}{a^2},\label{TLp}\end{eqnarray}
 where
 \begin{eqnarray} 
   {\cal B}_1(a)&=&\Omega_0^{(r)}+\Omega_0^{(m)}\,a+\Omega_0^K\,a^2+\Omega_0^\Lambda\,a^4,\nonumber\\
   {\cal B}_2(a)&=&-2\Omega_0^{(r)}-\Omega_0^{(m)}a+2\Omega_0^\Lambda a^4.\label{vardefs}
 \end{eqnarray}
 \item Eulerian frame $T_{\tiny{\hbox{E}}}^{ab}$ (CMB and Lambda):
 \begin{equation}\frac{8\pi\,\rho}{3 H_0^2}=\frac{\Omega_0^{(r)}}{a^4}+\Omega_0^\Lambda,\qquad \frac{8\pi\,p}{3 H_0^2}=\frac{\Omega_0^{(r)}}{3a^4}-\Omega_0^\Lambda.\label{CMBframe}\end{equation}
 \end{itemize}
Peculiar velocities can be computed from \eqref{pecvel}, which takes the form
\begin{equation} v^a = v \delta^a_r, \qquad v = v_{\tiny{\hbox{pec}}}= \frac{Q}{\hat \rho+\hat p}\approx  \frac{Q}{ \rho+p},\label{pecvel1}\end{equation}
where $\hat\rho$ and $\hat p$ are given  by  \eqref{TLrhop}.

\subsection{Scenario 2: inhomogeneous non-comoving dark energy} 

The terms $\Delta^{(p)}$ and $\Delta^{(\rho)}$ of order $\epsilon_0$ in \eqref{TLrhop} can also be used to describe an inhomogeneous dark energy source with the cosmological constant as the zero order term, leading to the following forms for the density and pressure 
\begin{equation}\frac{8\pi\,\hat\rho^{(de)}}{3 H_0^2}=\Omega_0^{\Lambda}+\epsilon_0\,\Delta^{(\rho)},\qquad \frac{8\pi\,\hat p^{(de)}}{3 H_0^2}=-\Omega_0^{\Lambda}+\epsilon_0\,\Delta^{(p)}.\label{inhomDE}\end{equation}
The Lagrangian frame $\hat T_{\tiny{\hbox{L}}}^{ab}$ corresponds to matter (baryons and CDM) as  homogeneous dust and DE as the inhomogeneous perfect fluid \eqref{inhomDE}, while the Eulerian frame $T_{\tiny{\hbox{E}}}^{ab}$ is the CMB radiation as a homogeneous perfect fluid (a photon gas).
\begin{eqnarray}
 \hat T_{\tiny{\hbox{L}}}^{ab} &=&( \hat\rho^{(m)}+  \hat\rho^{(de)}) \hat u^a\hat u^b +\hat p^{(de)} \hat h^{ab},\\
T_{\tiny{\hbox{E}}}^{ab}&=&\rho^{(r)} u^a u^b +p^{(r)} h^{ab},\qquad p^{(r)}=\frac13\rho^{(r)}, \end{eqnarray}
with the total energy momentum $T^{ab}=T_{\tiny{\hbox{L}}}^{ab}+T_{\tiny{\hbox{E}}}^{ab}$ in the Eulerian frame given by \eqref{TabE}.  The Lagrangian and Eulerian frames are
 \begin{itemize}
 \item Lagrangian frame $\hat T_{\tiny{\hbox{L}}}^{ab}$ (baryons, CDM and dark energy):
 \begin{eqnarray}\frac{8\pi\,\hat\rho}{3 H_0^2}&=&\frac{\Omega_0^{(m)}}{a^3}+\Omega_0^{(\Lambda)}+\epsilon_0\,\Delta^{(\rho)},\nonumber\\
  \frac{8\pi\,\hat p}{3 H_0^2}&=&-\Omega_0^{(\Lambda)}+\epsilon_0\,\Delta^{(p)},\label{TLrhop2}\end{eqnarray}
 where $\Delta^{(\rho)}$ and $\Delta^{(p)}$ are given by \eqref{TLrho} and \eqref{TLp}.
  \item Eulerian frame $T_{\tiny{\hbox{E}}}^{ab}$ (CMB):
 \begin{equation}\frac{8\pi\,\rho}{3 H_0^2}=\frac{\Omega_0^{(r)}}{a^4},\qquad \frac{8\pi\,p}{3 H_0^2}=\frac{\Omega_0^{(r)}}{3a^4},\label{CMBframe2}\end{equation}
 \end{itemize}
 with peculiar velocities given by \eqref{pecvel1} with $\hat\rho$ and $\hat p$ in \eqref{TLrhop2}.
 
 \subsection{Conservation}
 
 It is evident from the forms of $\rho$ and $p$ in \eqref{TLrhop}-\eqref{CMBframe}, \eqref{inhomDE} and \eqref{TLrhop2}-\eqref{CMBframe2} that individual fluids in these mixtures are not conserved.  The energy momentum current associated with the total energy momentum in the Eulerian frame (equation \eqref{Tab}) can be projected into $u^a$ and $h_{ab}$ as follows
 \begin{eqnarray}
 u_a{\cal J}^a &=& u_a\nabla_b T_{\tiny{\hbox{total}}}^{ab}=\dot \rho+(\rho+p)\Theta +\left(\ln \frac{af^2}{(1+\epsilon_0 bF)^3}\right)_{,r}Q\nonumber\\
 +Q_{,r},\label{uJ}\\
 h_{ab}{\cal J}^b&=&h_{ab}\nabla_c T_{\tiny{\hbox{total}}}^{bc}=\left[\frac35\Theta Q+\dot Q+\frac{(1+\epsilon_0 bF)^2}{a}\,p_{,r}\right]\,\delta_a^r,\nonumber\\
 \label{hJ}
 \end{eqnarray}
 where $\rho$ and $p$ are the total density and pressure and $Q$ and $\Theta$ are given by \eqref{eqQCF2} and \eqref{eqThCF2}.  We examine conservation in the Eulerian frame for Scenario 1 (Scenario 2 is analogous), by assuming $\hat\rho\approx \rho$ and $\hat p\approx p$ to rewrite the total energy momentum $T_{\tiny{\hbox{total}}}^{ab}=T^{ab}_{\tiny{\hbox{L}}}+T^{ab}_{\tiny{\hbox{E}}}$ in \eqref{TabE}-\eqref{TabE2} as
 \begin{eqnarray}
 T_{\tiny{\hbox{L}}}^{ab}&=&(\rho_{(0)}^{(m)}+\epsilon_0\,\delta^{(\rho)})u^au^b+\epsilon_0\,\delta^{(p)}h^{ab}+2q^{(a}u^{b)},\label{TabL1}\\
 T_{\tiny{\hbox{E}}}^{ab}&=&(\rho_{(0)}^{(r)}+\Lambda)u^au^b+(p_{(0)}^{(r)}-\Lambda)h^{ab}),\label{TabE1}
 \end{eqnarray}
 where $\rho_{(0)}^{(m)},\,\rho_{(0)}^{(r)},\,p_{(0)}^{(r)}=\rho_{(0)}^{(r)}/3$ depend only on time and $Q$ is related to the peculiar velocity as 
 \begin{equation} q^a =Q\delta^a_r= [\rho_{(0)}^{(m)}+\epsilon_0\,(\delta^{(\rho)}+\delta^{(p)})] v\,\delta^a_r.\end{equation}
 It is straightforward to show from \eqref{TabE2} and \eqref{TabE1} that the spacelike projection is conserved: $h_{ab}{\cal J}^b=0$ in \eqref{hJ} for $T_{\tiny{\hbox{total}}}^{bc}$, since  $h_{ab}{\cal J}^b=0$  holds trivially for the photon fluid with $T_{\tiny{\hbox{E}}}^{ab}$ that does not contain $Q$ and whose pressure gradient is zero. As a consequence, $h_{ab}\nabla_c T_{\tiny{\hbox{total}}}^{bc}= h_{ab}\nabla_c T_{\tiny{\hbox{L}}}^{bc}=0$ holds.  However, while the projection $u_a{\cal J}^a$ is conserved for $T_{\tiny{\hbox{total}}}^{ab}$, it is not conserved for the individual fluids:
 \begin{eqnarray}    u_a\nabla_b T_{\tiny{\hbox{E}}}^{ab}&=&\dot \rho_{(0)}^{(r)}+\frac43\rho_{(0)}^{(r)}\Theta=\dot \rho_{(0)}^{(r)}+4\rho_{(0)}^{(r)}\left(\frac{\dot a}{a}-\frac{\epsilon_0 \dot b F}{1+\epsilon_0 b F}\right)\nonumber\\
 &=&\frac{4\epsilon_0 \rho_{(0)}^{(r)}\dot b F}{1+\epsilon_0 b F}=\Gamma\ne 0,\\
  u_a\nabla_b T_{\tiny{\hbox{L}}}^{ab}&=&\dot \rho_{(0)}^{(m)}+\rho_{(0)}^{(m)}\Theta +\epsilon_0\,[\dot\delta^{(\rho)}+(\delta^{(\rho)}+\delta^{(p)})]\Theta\nonumber\\
  && +\left(\ln \frac{af^2}{(1+\epsilon_0 bF)^3}\right)_{,r}Q+Q_{,r}\nonumber\\
   &=&\frac{4\epsilon_0 \rho_{(0)}^{(m)}\dot b F}{1+\epsilon_0 b F}+\epsilon_0\,[\dot\delta^{(\rho)}+(\delta^{(\rho)}+\delta^{(p)})]\Theta\nonumber\\
   && +\left(\ln \frac{af^2}{(1+\epsilon_0 bF)^3}\right)_{,r}Q+Q_{,r}=-\Gamma.\nonumber\\ \end{eqnarray}
 where $\Gamma\ne 0$ is the interaction term that should be determined on physical assumptions, a task that is beyond the scope of this paper.  
 
\section{Numerical results}\label{sec:numres}

The models under consideration have positive spatial curvature, but current observations favor spatial flatness. Therefore, we choose the free parameters $\Omega_0^{(r)}, \Omega_0^{(m)}$ and $ \Omega_0^{\tiny{\Lambda}}$ so as to obtain a very small positive spatial curvature, close to spatial flatness
\begin{eqnarray}    \Omega_0^{(r)}&=&5\times 10^{-5},\quad\Omega_0^{(m)}=0.31,\quad \Omega_0^{\tiny{\Lambda}}=0.7,\nonumber\\
\Omega_0^{\tiny{\hbox{K}}}&=&\Omega_0^{(r)}+\Omega_0^{(m)}+\Omega_0^{\tiny{\Lambda}}-1=0.01,\label{Omegapars}\end{eqnarray}
providing a reasonably good approximation to the time evolution of a spatially flat $\Lambda$CDM model (though as mentioned before, the FLRW equations in \eqref{eqrhoCF2}-\eqref{eqpCF2}  do not determine the time evolution even at the symmetry centers). 

The two scenarios described previously require verifying the signs of the terms $\Delta^{(\rho)}$ and $\Delta^{(p)}$ at order $\epsilon_0$. The inhomogeneous dark matter fluid in Scenario 1 necessarily requires (from \eqref{TLrhop})  $\Delta^{(p)}>0$, while a negative pressure  $\Delta^{(p)}<0$ is appropriate for the inhomogeneous dark energy fluid in \eqref{TLrhop2}. To verify the signs of  $\Delta^{(\rho)}$ and $\Delta^{(p)}$ we compute  their expansion around $a=0$
\begin{eqnarray}    \Delta^{(\rho)} &\approx& 2\epsilon_0\left(1+\frac{2\Omega_0^{(r)}F}{1+\epsilon_0F}\right)\frac{1}{a^2},\nonumber\\
 \Delta^{(p)} &\approx&-\frac{2\epsilon_0}{3}\left(1+\frac{6\Omega_0^{(r)}F }{1+\epsilon_0F}\right)\frac{1}{a},\end{eqnarray}
showing that the fact that $\Omega_0^{(r)} \ll 1$ implies $ \hbox{sign} (\Delta^{(\rho)})=\hbox{sign} (\epsilon_0)$, while $ \hbox{sign} (\Delta^{(p)})=-\hbox{sign} (\epsilon_0)$.  Therefore, we need $\epsilon_0<0$ to get $\Delta^{(p)}>0$ required for the pressure of baryons and CDM in Scenario 1, which also implies $\Delta^{(\rho)}<0$, but the whole matter density $\Omega_0^{(m)}/a^3-|\Delta^{(\rho)}|$ remains positive since $|\Delta^{(\rho)}|\ll \Omega_0^{(m)}/a^3$. For the dark energy fluid in Scenario 2 we expect $\rho>0$ and $p<0$, hence we use $\epsilon_0>0$.

\subsection{Peculiar velocities}

\begin{figure}[h]
\centering
\includegraphics[width=\linewidth]{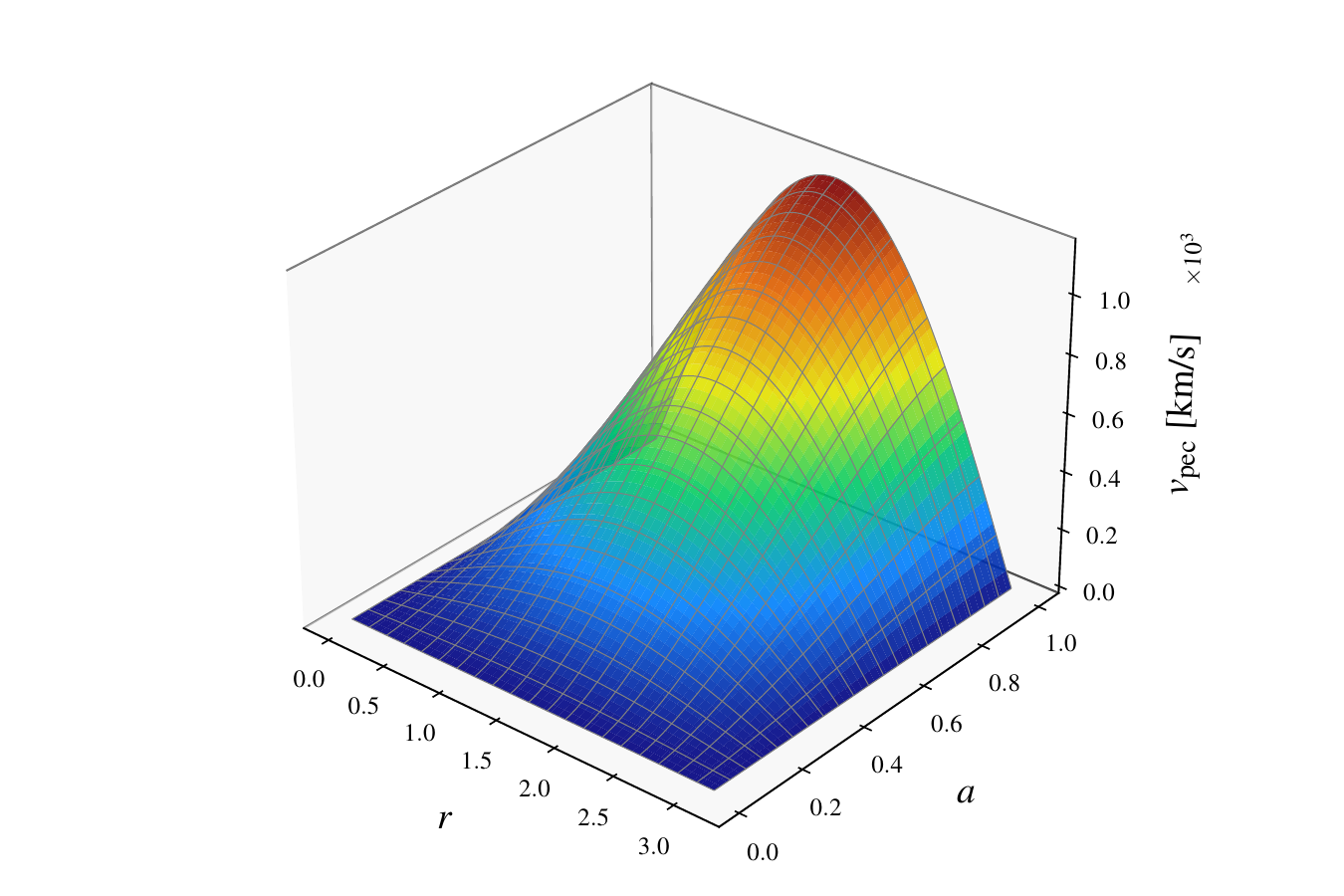}
\caption{Peculiar velocities, $v_{\tiny{\hbox{pec}}}$, as a function of $(a,r)$. Notice the peculiar velocity vanishes at $a=0$ for all values of $r$ and also vanishes at the symmetry centers $r=0,\pi$. Refer to main text for discussion.}
\label{fig:fig1}
\end{figure}

To examine \eqref{pecvel1}, we need to evaluate $Q$ from \eqref{eqQCF2} for $b$ given by \eqref{bfun} and $f=\sin r$:
\begin{equation}\frac{8\pi Q}{3H_0^2} = \frac{4\epsilon_0 \sqrt{{\cal B}_1(a)}\,\sin r}{3a^2},\end{equation}
where ${\cal B}_1(a)$ is given by \eqref{vardefs}. Since $ \hbox{sign} (Q)=\hbox{sign} (\epsilon_0)$,  the quotient \eqref{pecvel1} is necessarily negative for Scenario 1 and positive for Scenario 2. Considering Scenario 2 first, we plot in Fig.~\ref{fig:fig1} the peculiar velocity $v_{\tiny{\hbox{pec}}}$ from \eqref{pecvel1} for the parameters in \eqref{Omegapars} and $\epsilon_0=0.005$. The graph shows that $v_{\tiny{\hbox{pec}}}=0$ at the initial singularity $a=0$ and at both symmetry centers $r=0,\,\pi$.\\

%\noindent
%Place Figure 1\\ 

%
%\noindent
This is a physically consistent result, since peculiar velocities are not expected to occur in the early Universe (close to $a=0$) and also not in the symmetry centers, which do not admit privileged directions for being worldlines of locally isotropic observers.  As the graph in Fig.~\ref{fig:fig1} shows, peculiar velocities constitute a radial field that for each fixed time (fixed $a$) varies with the radial coordinate $r$ reaching a maximum at $r=\pi/2$. For present cosmic time $a=1$ radial peculiar velocities vary in the range $0\leq v_{\tiny{\hbox{pec}}}\leq 2400$ km/s, which is the range of values of peculiar velocities reported for galactic clusters and superclusters \cite{sankhyayan2023identification,kopylova2014peculiar,kopylova2024fundamental} and in the CMB dipole  \cite{wagenveld2023cosmic, singal2011large}.

\begin{figure}[h]
\centering
\includegraphics[width=\linewidth]{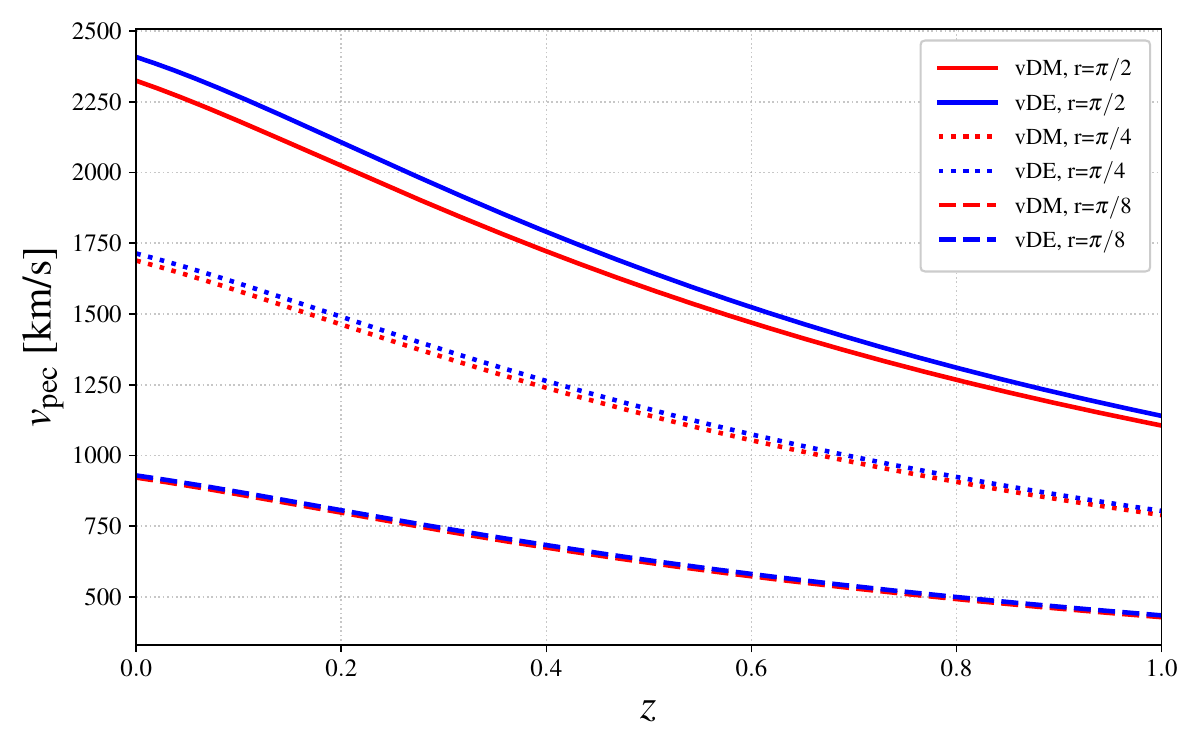}
\caption{Peculiar velocities, $v_{\tiny{\hbox{pec}}}$, from Scenario 1 (continuous curves) and Scenario 2 (dashed curves) as  functions of redshift $z=1/a-1$ for various values of $r$, corresponding to dark matter (Scenario 1) and dark energy (scenario 2). Notice that $v_{\tiny{\hbox{pec}}}$ is slightly larger in Scenario 2.}
\label{fig:fig2}
\end{figure}

For Scenario 1 peculiar velocities from \eqref{pecvel1} with same parameters as in Scenario 2 leads to $v_{\tiny{\hbox{pec}}}\leq 0$. The positive/negative sign of the radial component $v_{\tiny{\hbox{pec}}}$ of the velocity field $v^a=v_{\tiny{\hbox{pec}}}\delta^a_r$ simply indicates that $v_{\tiny{\hbox{pec}}}$ increases/decreases with increasing $r$. Hence, for $v_{\tiny{\hbox{pec}}}>0$ the velocity field flows from the symmetry center $r=0$ towards the other center $r=\pi$, with $v_{\tiny{\hbox{pec}}}<0$ indicating the reverse flow. However, in Scenario 1 $|v_{\tiny{\hbox{pec}}}|$ has almost the same numerical values as $v_{\tiny{\hbox{pec}}}>0$ shown in Fig.~\ref{fig:fig1}. This is expected, since a change of sign in $\epsilon_0$ has a negligible effect because we assume $|\epsilon_0|\ll 1$. For both scenarios the numerator $Q$ of \eqref{pecvel1} has the same magnitude with different sign, while regarding the denominator we have very similar values (the $\Lambda$ term cancels in $\rho+p$): 
\begin{eqnarray} \rho+p=\frac{\Omega_0^{(m)}}{a^3}-|\epsilon_0|\,[\Delta^{(\rho)}+\Delta^{(p)}],\qquad \textrm{Scenario\, 1},\label{Sc1}\\
\rho+p=\frac{\Omega_0^{(m)}}{a^3}+\epsilon_0[\Delta^{(\rho)}+\Delta^{(p)}], \qquad \textrm{Scenario \,2}. \label{Sc2}\end{eqnarray}
To examine the difference between $|v_{\tiny{\hbox{pec}}}|$ from \eqref{Sc1} and $v_{\tiny{\hbox{pec}}}$ from \eqref{Sc2}, we plot them in Fig.~\ref{fig:fig2} for the same parameters as in Fig.~\ref{fig:fig1} for radial value $r=\pi/2$ of maximal growth of $v_{\tiny{\hbox{pec}}}$. As shown in the figure $|v_{\tiny{\hbox{pec}}}|=2400$ km/s in \eqref{Sc1}, about 10\% larger than $v_{\tiny{\hbox{pec}}}=2240$ km/s in  \eqref{Sc2}.
%\noindent
%Place Figure 2\\
%\noindent
%Place Figure 1\\

\subsection{Physical properties of the fluids}

\subsubsection{Inhomogeneous matter (baryons and CDM).}

A pure dust energy momentum tensor is an accurate approximation to CDM \cite{armendariz2014cold}, but a small nonzero pressure might be justified at order $\epsilon_0$ in \eqref{TLp} since we are assuming in \eqref{Omegapars} that $\Omega_0^{(m)}=0.31$ represents CDM and baryons. However, assuming that the mixture of baryons and CDM is a non-collisional ensemble of particles, their pressure should roughly take the form $p\sim \rho \,v_d^2$, with $v_d$ a dispersion velocity that  must be consistent with characteristic velocities of self-gravitating structures which range up to $\sim 2000$ km/s in galactic clusters and superclusters \cite{sankhyayan2023identification,kopylova2014peculiar,kopylova2024fundamental} and in the CMB dipole  \cite{wagenveld2023cosmic,singal2011large}. 

Since dispersion velocities should be of the same order of magnitude as peculiar velocities, we examine the relation $v_d/c=\sqrt{p/\rho}$ for Scenario 1 in \eqref{TLrhop}-\eqref{TLp} for the parameters used in Fig.~\ref{fig:fig1}, with $\epsilon_0=-0.005$ that yield realistic peculiar velocities up to 2400 km/s from \eqref{pecvel1}. We obtain at present time $a=1$ the value $\sqrt{p/\rho}= 0.16$, which implies characteristic velocities of 47,400 km/s that are wholly unrealistic. To obtain a ratio $v_d\sim 2000$ km/s compatible with the peculiar velocities in Figs.~\ref{fig:fig1} and \ref{fig:fig2}, we need to set $\epsilon_0=-0.00001$, which practically reduces the perfect fluid \eqref{TLrhop} to pure dust with peculiar velocities \eqref{pecvel1} to $v_d\sim 5$ km/s consistent with dispersion velocities of CDM \cite{armendariz2014cold}. We plot in Fig.~\ref{fig:fig3} $\Delta^{(\rho)}$ and $|\Delta^{(p)}|$ from \eqref{TLrhop} for $\epsilon_0=-0.005$ and $\epsilon_0=-0.00001$\\ 

\noindent

\begin{figure}[h]
\centering
\includegraphics[width=0.8\linewidth]{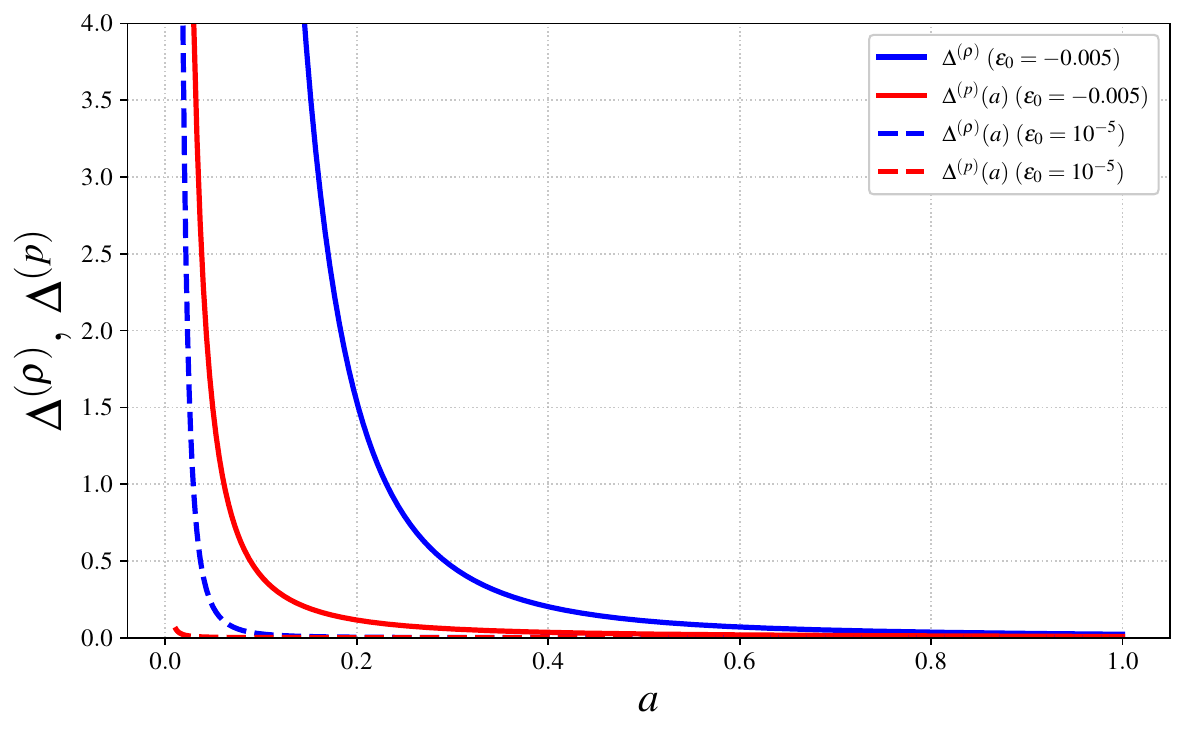}
\caption{%
$\Delta^{(\rho)}$ and $|\Delta^{(p)}|$ as functions of $a$. Continuous curves depict $\epsilon_0=-0.005$, while dashed curves show $\epsilon_0=-0.00001$. See main text for details.}
\label{fig:fig3}
\end{figure}
%Place Figure 3\\
\iffalse
\begin{figure}[h]
\centering
\includegraphics[width=0.8\linewidth]{PlotsShearSimp/Drho_Dp_vs_a.pdf}

\vspace{0.5cm}

\includegraphics[width=0.8\linewidth]{PlotsShearSimp/Dp_over_Drho_vs_a.pdf}

\caption{%
(Top) $\Delta \rho$ and $|\Delta p|$ as functions of $a$. Continuous curves depict $\epsilon_0=-0.005$, while dashed curves show $\epsilon_0=-0.00001$. 
(Bottom) Description of the second plot.}
\label{fig:fig3}
\end{figure}
\fi

\noindent
 Another potential problem with the physical interpretation of the perfect fluid of Scenario 1  is the fact that $p>0$ in \eqref{TLrhop} requires $\epsilon_0<0$ which implies $\Delta^{(\rho)}<0$, though the total density remains positive because $\Omega_0^{(m)}/a^3\gg |\Delta^{(\rho)}|$.  Conversely,  $\epsilon_0>0$ leads to $\Delta^{(\rho)}>0$ but $\Delta^{(p)}<0$. Therefore, $\Delta^{(\rho)}$ and $\Delta^{(p)}$ cannot play the role of internal energy analogues in \eqref{TLrhop}-\eqref{TLp} in this perfect fluid.  
 
 For baryons and CDM described as particle ensembles in Scenario 1 $\rho$ and $p$ lead to reasonably realistic   peculiar velocities from \eqref{pecvel1} (Fig.~\ref{fig:fig2}), but predict unrealistic dispersion velocities. However, if CDM is described as a scalar field (as Bose-Einstein condensate), then peculiar velocities need not be consistent with the dispersion ratio $v_d/c=\sqrt{p/\rho}$ and pressure can even be negative \cite{chavanis2025review}, though exploring this possibility is outside the scope of this paper.  
 
 \subsubsection{Varying dark energy}    

In Scenario 2 we assume the varying perfect fluid dark energy given in  \eqref{inhomDE}, with $\Lambda$ the term of zero order in $\epsilon_0$.  As shown in Fig.~\ref{fig:fig2}, we obtain for this scenario and same free parameters peculiar velocities of up to 2400 km/s (slightly larger than in Scenario 1). Given the lack of knowledge on the fundamental nature of dark energy, the  plausibility of peculiar velocities in Scenario 2 need not be restricted by their compatibility of the dark energy source \eqref{inhomDE} with the dispersion ratio $v_d/c=\sqrt{p/\rho}$ (as in Scenario 1). 

\begin{figure}[h]
\centering
\includegraphics[width=0.8\linewidth]{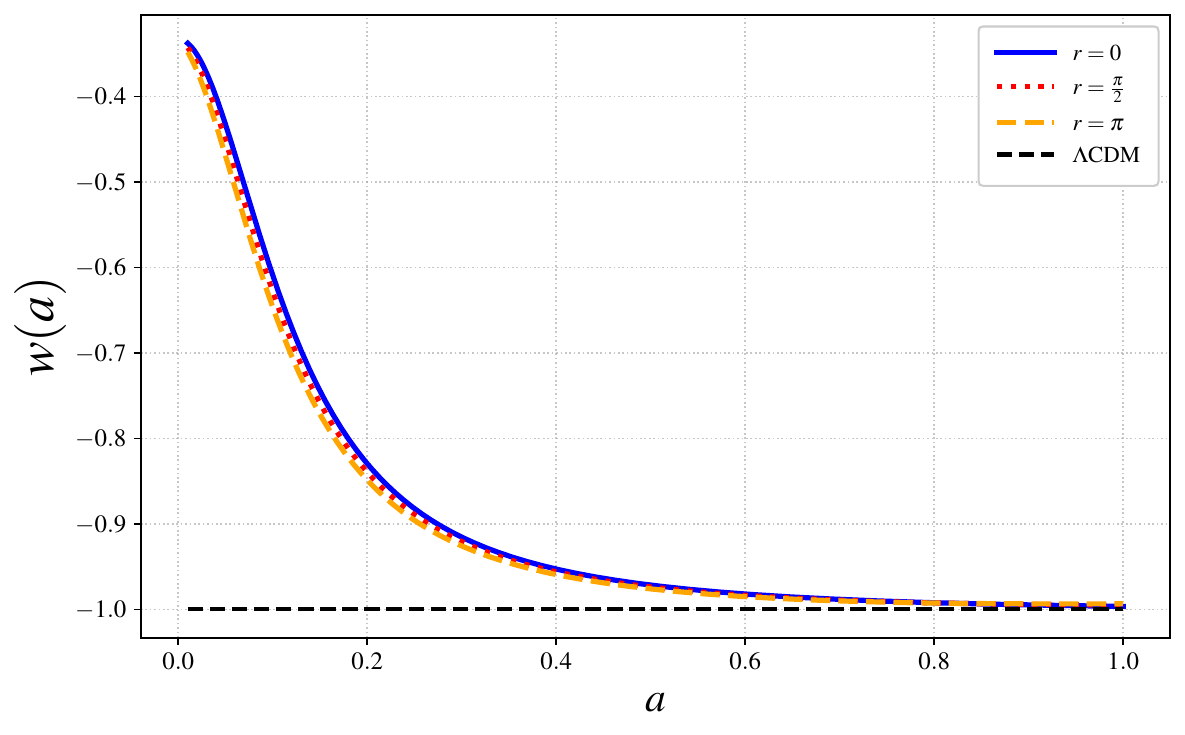}

\vspace{0.5cm}

\includegraphics[width=0.8\linewidth]{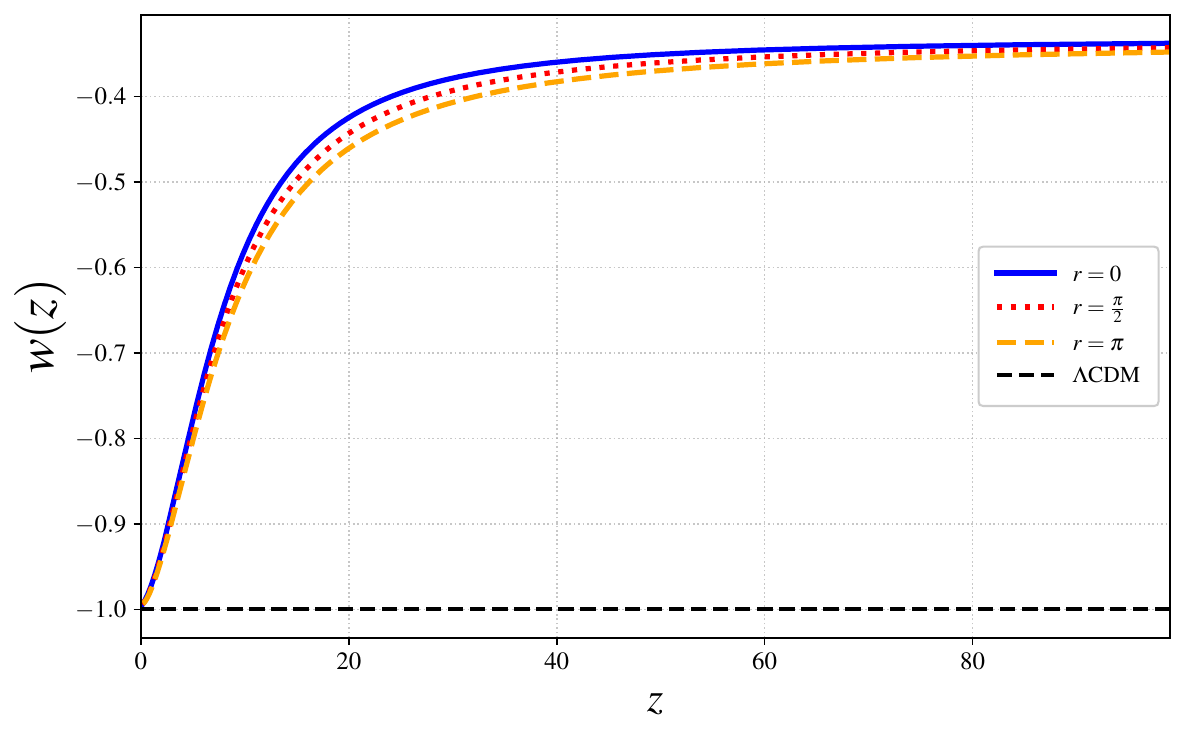}

\caption{%
Ratio $w=p^{(DE)}/\rho^{(DE)}$. 
Top: $w$ as a function of $a$. 
Bottom: $w$ as a function of $z$ for various values of $r$. Details are discussed in the text.}
\label{fig:fig4}
\end{figure}

While  $\epsilon_0>0$ assures a negative total dark energy pressure, it is important to verify the evolution of the ratio $w=p^{(de)}/\rho^{(de)}$ and find if it approaches $w\to -1$  in present cosmic time $a=1$. We plot in Fig.~\ref{fig:fig4} the ratio $w$ for the perfect fluid \eqref{inhomDE} as a function of $a$ (Top panel of Fig.~\ref{fig:fig4}) and of redshift (Bottom pannel of Fig.~\ref{fig:fig4}) for the parameters used for the peculiar velocities from \eqref{pecvel1} for Scenario 2 in Fig.~\ref{fig:fig2}. \\ 

\noindent

%Place Figure 4\\

\noindent
As shown in Fig.~\ref{fig:fig4} (Top panel), $w$ evolves from $w=-1/3$ at $a=0$, reaching (as expected) values very close to the cosmological constant $w=-1$ at present day $a=1$. The figure shows how $w$ is almost a purely time dependent function of $a$, with almost negligible spatial variation. Bottom panel of Fig.~\ref{fig:fig4} depicts the radial profile of $w$ at $a=0$ and at different redshifts (computed as $1+z=1/a$), showing how the spatial variation is negligible and the approach $w\to -1 $ occurs faster at the symmetry center $r=\pi$.      

The fact that $w=-1/3$ at $a=0$ and $w\to -1$ asymptotically denotes a strange transition of dark energy from a cosmic string state to a de Sitter state. However, the models we are considering are only expected to be valid in late cosmic times, for which $w\to -1$ is a satisfactory result. A varying dark energy is currently the subject of an abundant literature exploring a variety of different models (k-essence, Chaplygin gas) \cite{o2021can,keshav2024interacting} and subjecting their predictions to current observational constraints  \cite{chakraborty2025desi,notari2024consistent}. Relating this research to the dark energy fluid in  \eqref{inhomDE} is beyond the scope of this paper.  

\subsection{Redshifts}

Peculiar velocities can be estimated from the difference between the observed redshift from a source and the recession velocity due to the Hubble flow, which we can assume to be associated with an idealized FLRW spacetime that can be identified with the CMB frame \cite{davis2014deriving,boruah2020cosmic}. We examine the relation between these redshifts for the models under consideration. 

\begin{figure}[h]
\centering
\includegraphics[width=0.8\linewidth]{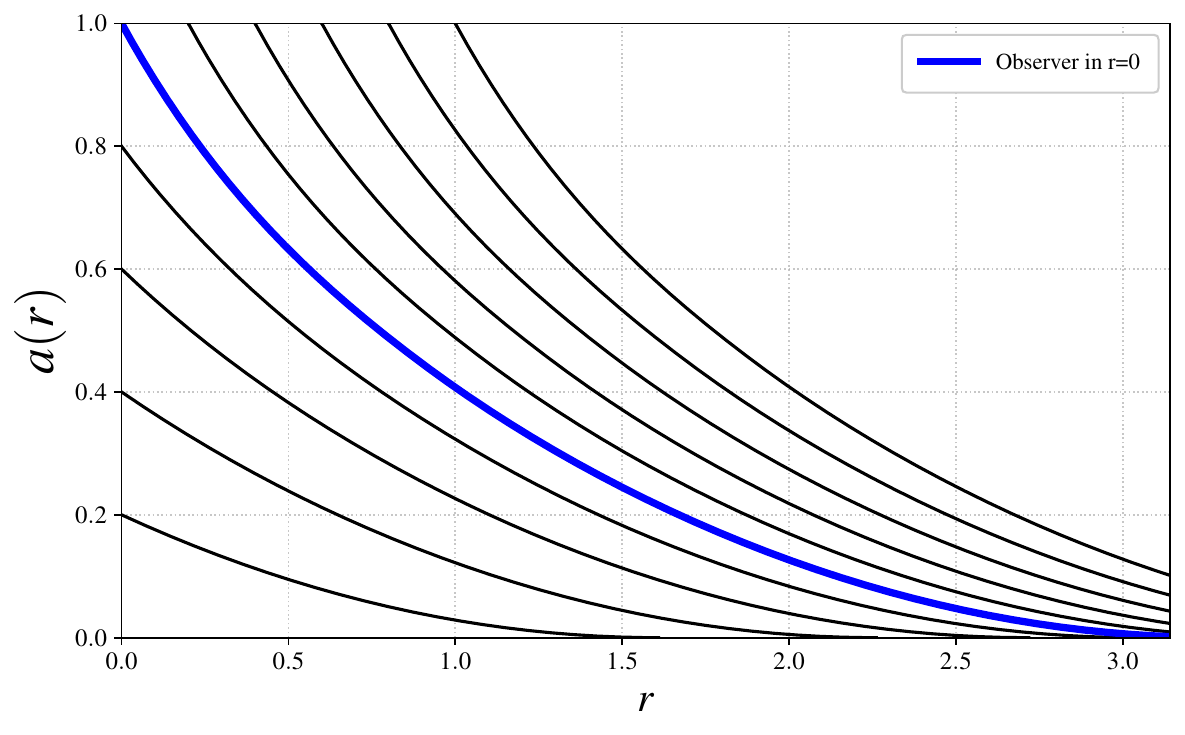}

\vspace{0.5cm}

\includegraphics[width=0.8\linewidth]{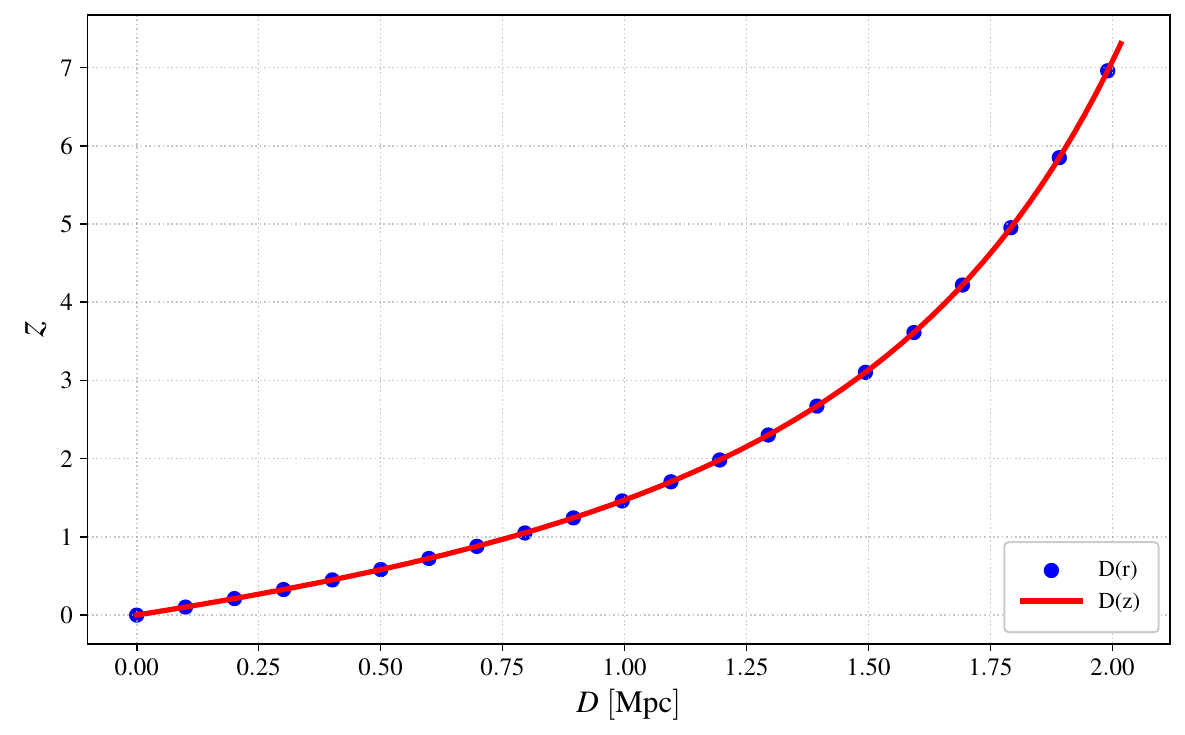}

\caption{%
Top: Ingoing null geodesic, highlighted curve shows the observer at $r=0$.  
and $a=1$, Bottom: Redshift $z$ and $r$ as functions of the comoving distance $D$. Refer to main text for further discussion.}
\label{fig:fig5}
\end{figure}

Observers receive emitted light rays parametrized by an affine parameter $x^a(\nu)$,  whose tangent vector $k^a=dx^a/d\nu$ is null ($k_ak^a=0$) and satisfies the null geodesic equation $k^b\nabla_b   k^a=0$. Since the models are spherically symmetric, we can consider radial null geodesics $x^a=[t(\nu),\,r(\nu),0,0]$ with tangent vector $k^a = [k^t(\nu),\,k^r(\nu),0,0] = [\mu_0 a/L,1,0,0]$, where $\mu_0=\pm 1$ and we used $k_ak^a=0$. The null geodesic equation is:
\begin{eqnarray}\frac{dt}{d\nu}&=&\frac{a\left[-\epsilon_0^2\mu_0\,f\, F e^{-2a^2}+\epsilon_0\left((2a^2+1)\dot a\,F -\mu_0 f\right)e^{-a^2}\right)+2\,a\,\dot a]}{\left(1+\epsilon_0 e^{-a^2}\,F\right)^3},\nonumber\\
\label{geod1}\\
\frac{dr}{d\nu}&=& \frac{
a\left[
-\epsilon_0^2\,\mu_0\,F\,f\,e^{-2a^2}+\epsilon_0
\left(\mu_0(2a^2+1)\dot a\,F - f
\right)e^{-a^2}+2\mu_0\dot a\right]}{\left(1+\epsilon_0 e^{-a^2}\,F\right)^2},\nonumber\\
\label{geod2}\end{eqnarray}
where $f,\,F$ and $\dot a$ are defined in \eqref{Fchi2} and \eqref{zeroO}. Using the chain rule $da/d\nu=(da/dt)( dt/d\nu)$ to form $[da/dr]_{\tiny{\hbox{null}}}$ along the null geodesic and expanding around $\epsilon_0$ we obtain the following equation 
\begin{equation} \left[\frac{da}{dr}\right]_{\textrm{null}}=\mu_0\,\dot a\, \left(1-\epsilon_0\,a^{-a^2}\,F\right),\label{geod3}\end{equation}
which, as we show in  \ref{sec:series}, provides solutions that are a very accurate approximation of the exact equation (see exact equation in \eqref{nullgeod4}). We choose an observer  receiving light signals along the symmetry center $r=0$ at present cosmic time $a=1$, hence $\mu_0=-1$ (ingoing geodesics from emitters at $r>0$). Since $z=0$ and $a=1$ at the observation point, the redshift formula becomes 
\begin{equation} 1+z =\frac{\left(u_a\,k^a\right)_{\textrm{ob}}}{\left(u_a\,k^a\right)_{\textrm{em}}}=\frac{1+2\epsilon_0e^{-a^2}\,F}{a},\label{redshift}\end{equation}
where $a=a(r)$ along the null geodesic must satisfy  \eqref{geod3}.  We solve numerically equation \eqref{geod3} for the parameters in \eqref{Omegapars} (omitting the radiation term: $\Omega_0^{(r)}=0$) and $\epsilon_0=0.005$ used in the previous  section. The FLRW particular case follows readily  by setting $\epsilon_0=0$ in \eqref{geod3} and \eqref{redshift}, so that for $\epsilon_0=0.005\ll 1$ and a very small spatial curvature $\Omega_0^{(K)}=0.01$, the solutions are almost indistinguishable from the spatially flat FLRW solutions. Hence, we compute the  comoving distance as a function of $z$ using the FLRW formula: 
\begin{equation}D(z)=\frac{c}{H_0}
\int_0^z{\frac{dz}{\sqrt{\Omega_0^{(m)}(1+z)^3+\Omega_0^{(K)}(1+z)^2+\Omega_0^{(\Lambda)}}}},\label{Dz} \end{equation}
Top panel of Fig.~\ref{fig:fig5} displays radial ingoing null geodesics, highlighting the geodesic reaching the observer at $r=0,\,a=1$, while bottom panel of Fig.~\ref{fig:fig5} depicts the redshift \eqref{redshift} as a function of the comoving distance \eqref{Dz}.\\

\noindent

%Place Figure 5a and Figure 5b\\

\begin{figure}[h]
\centering
\includegraphics[width=\linewidth]{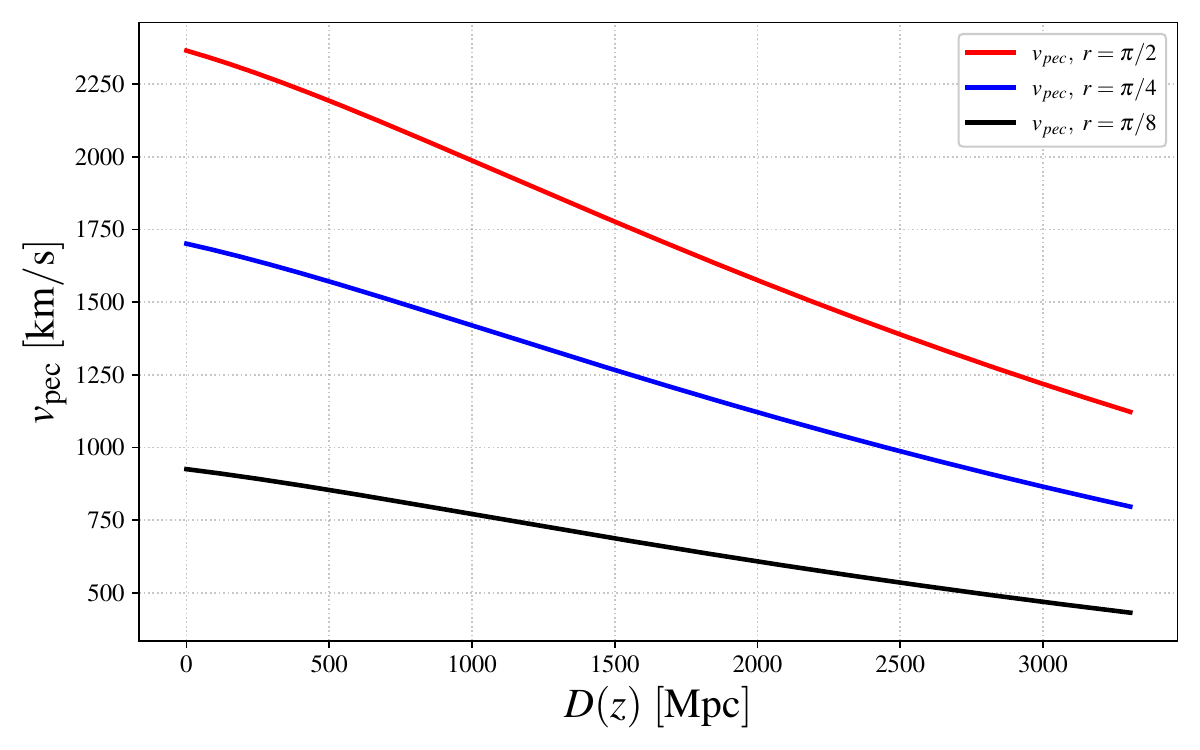}
\caption{Peculiar velocities obtained from the redshift relation \eqref{vp} as a function of the comoving distance $D$. Notice the values of these velocities are consistent with values displayed in Figures \ref{fig:fig1} and \ref{fig:fig2} and with ranges reported in the literature.}
\label{fig:fig6}
\end{figure}

\noindent
Peculiar velocities $v_{\tiny{\hbox{pec}}}$ can be estimated from the observed redshift $z$ and the redshift associated with the Hubble flow of a background FLRW model $\bar z$. Assuming $z$ given by  \eqref{redshift} and $\bar z$ associated with the nearly spatially flat FLRW with parameters \eqref{Omegapars}, peculiar velocities follow as \cite{davis2014deriving,boruah2020cosmic}
\begin{equation}v_{\tiny{\hbox{pec}}} = \frac{c(z-\bar z)}{1+\bar z}=\frac{\epsilon_0\, e^{-a^2}(1-\cos r)\,c}{1-a},\label{vp}\end{equation}
where $a=a(r)$ is a solution of \eqref{geod3}. As shown in Fig.~\ref{fig:fig6} displaying $v_{\tiny{\hbox{pec}}}$ as a function of the comoving distance \eqref{Dz} for the parameters \eqref{Omegapars} with $\epsilon_0=0.005$, the peculiar velocities increase with increasing $D(z)$ in the range up to 2000 km/s (as reported in the literature \cite{wagenveld2023cosmic,sankhyayan2023identification,kopylova2014peculiar,kopylova2024fundamental,singal2011large}), which is of the same order of magnitude as peculiar velocities shown in Figs.~\ref{fig:fig1} and \ref{fig:fig2}. 
\\

\noindent

%Place Figure 6\\

\noindent  

\section{Discussion and Conclusion}\label{sec:conc}

The purpose of this article is to present and discuss the potential of generating non-trivial peculiar velocity fields through exact solutions of Einstein's equations whose source is an irrotational shear-free fluid with nonzero energy flux.  These peculiar velocities emerge by considering the energy momentum tensor of these spacetimes as mixtures of perfect fluids with different 4-velocities, with peculiar velocities emerging from the non-relativistic limit of the Lorentzian boost between the 4-velocities. In the metric based approach of linear cosmological perturbations \cite{ma1995cosmological,burgazli2020effect,cembranos2019non} peculiar velocities appear as relative velocities of matter-energy sources in the Newtonian longitudinal gauge, characterized by irrotational shear-free congruences. In the covariant relativistic approach \cite{tsagas2025large,maartens1998covariant,ellis2001general} to cosmological perturbations this gauge represents a quasi-Newtonian frame with an irrotational shear-free 4 velocity in an Eulerian frame, while matter-energy sources  evolve in a Lagrangian frame with a different 4-velocity. The novel result in this paper is to discuss considering a non-perturbative approach based on analytic solutions of Einstein's equations whose source is an irrotational shear-free fluid.  

We have presented the most general known solution that admits no isometries and examined in detail a simple particular case with spherical symmetry, conformal flatness, zero 4-acceleration and positive spatial curvature. This solution can be described as a nearly homogeneous model whose source is a fluid mixture of radiation, matter (baryons and CDM) and dark energy. We examined two fluid scenarios in which the fluid in the Eulerian frame is a photon CMB gas, placing in the Lagrangian frame a perfect fluid describing baryonic and dark matter (Scenario 1) and a perfect fluid describing a varying dark energy that tends asymptotically to a de Sitter state (Scenario 2). In both scenarios we were able to obtain peculiar velocities up to 2400 km/s that are consistent with observations on the CMB dipole \cite{wagenveld2023cosmic,sankhyayan2023identification,kopylova2014peculiar,kopylova2024fundamental,singal2011large}. 

Evidently,  the simple particular solution that we considered as an illustrative example has very limited dynamical freedoms, so it is not a realistic cosmological model, though it is not entirely unrealistic. While the pressure in the perfect fluid in Scenario 1 is positive and much smaller than the energy density, it is not sufficiently small to predict characteristic dispersion velocities at the present cosmic time that are compatible with the peculiar velocities obtained for the same model. Likewise, the equation of state $w$ (ratio of pressure to density) of the varying dark energy in Scenario 2 smoothly evolves from $w=-1/3$ at the Big Bang to almost $w=-1$ at the present cosmic time. However, it was not our purpose to derive a realistic model, but to use it to illustrate the potential of analytic solutions of Einstein's equations for irrotational shear-free fluids to generate peculiar velocity fields. We believe that more general solutions of this class, with more dynamical degrees of freedom, might allow modeling more elaborate peculiar velocity fields that can be useful tools in cosmological research.      

\begin{acknowledgements}
SN acknowledges financial support from SECIHTI postdoctoral grants program.
\end{acknowledgements}

\begin{appendix}
\section{Spherically symmetric models: regularity conditions and coordinate transformations}\label{sec:coords}

The conformally flat, non-accelerating and spherically symmetric sub-case of the metric  \eqref{metric1} in spherical coordinates is 
\begin{equation}  ds^2 = -dt^2 +\frac{dr^2+r^2(d\theta^2+\sin^2\theta d\phi^2)}{L^2},\qquad L=\lambda_1+\lambda_2\Phi,\label{metricCF1}\end{equation}
where $\Phi=\alpha + \gamma r^2$ with $r^2=\delta_{ij}x^ix^j$.  Since the functions $\lambda_1,\,\lambda_2$ in  \eqref{metricCF1} are arbitrary and $L$ is a solution of \eqref{2linearCF} ($d^2L/d\Phi^2=0$), we can rearrange these 2 functions to rewrite  \eqref{metricCF1} as
\begin{eqnarray} ds^2 &=& -dt^2 + \frac{a^2(t)\left[dr^2+r^2(d\theta^2+\sin^2\theta d\phi^2)\right]}{L^2},\nonumber\\
 L&=& 1+\epsilon_0 b(t) r^2/2,\label{metricSS1}\end{eqnarray}
where $a(t)$ and $b(t)$ are arbitrary functions and $\epsilon_0$ is an arbitrary constant. Since the  hypersurfaces orthogonal to $u^a=\delta^a_t$ (rest  frames) are conformally flat, a direct generalization of \eqref{metricSS1} is to consider rest frames that are  conformal to 3 dimensional spaces with constant nonzero curvature, leading to the metric
\begin{eqnarray}  ds^2 &=& -dt^2 + \frac{a^2(t)\left[dr^2+r^2(d\theta^2+\sin^2\theta d\phi^2)\right]}{\left(1+\frac14 k_0 r^2\right)^2L^2},\nonumber\\
  L&=& 1+\epsilon_0 b(t) \tilde F(r),\label{metricSS2}\end{eqnarray}
where $k_0$ is a constant with inverse square length units whose sign denotes the sign of the 3 dimensional spaces with constant curvature (it is not necessarily the sign of  the spatial part of the 4-dimensional metric, it is identified as $k_0=H_0^2\Omega_0^{(K)}$  in \eqref{zeroO}). The metric coefficient $L$ complying with \eqref{2linearCF} is the same as in \eqref{metricCF1}, but $\Phi$ must be modified as  
\begin{equation}\Phi(r)= \lambda_1 + \lambda_2 \tilde F(r),\qquad \tilde F(r)=\frac{r^2/2}{1+\frac14 k_0r^2}. \label{PhiF}\end{equation}
To obtain the metric \eqref{metricCF2} from \eqref{metricSS2} we apply the following coordinate transformation 
\begin{eqnarray}\bar r = \int_0^r{\frac{dr}{1+\frac14k_0\ell^2r^2}} = \left\{  \begin{array}{ll} 
          \ell\,r & k_0=0 \\
          2\arctan (\ell\,r/2)& k_0=1 \\
          2\,\hbox{arctanh}(\ell\,r/2) & k_0=-1 
    \end{array}\right.,\label{chi}\end{eqnarray}
where $\ell$ is a characteristic inverse length, leading to
\begin{eqnarray}    ds^2 &=& -dt^2 +\frac{a^2(t)\left[d\bar r^2+f^2(\ell \bar r)(d\theta^2+\sin^2\theta d\phi^2)\right]}{L^2},\nonumber\\ 
L&=&1+\epsilon_0 b(t)F(\ell\bar r),\label{metricSS2b}\end{eqnarray}
where the functions $f$ and $F$ are
\begin{eqnarray}    f(\ell\bar r) &=&  \left\{  \begin{array}{ll} 
         \ell \bar r,& k_0=0,  \\
         \ell^{-1}\sin\,\ell \bar r,& k_0=1, \\
         \ell^{-1}\sinh\,\ell \bar r, & k_0=-1.  
       \end{array}\right.,
       \nonumber\\ 
      F(\ell \bar r)&=& \left\{  \begin{array}{ll} 
         \ell^2\bar r^2/2,& k_0=0,  \\
         1-\cos\,\ell\bar r,& k_0=1, \\
         \cosh\,\ell \bar r-1, & k_0=-1.  
       \end{array}\right.\label{fFchi}
\end{eqnarray}
The relation between $F(\ell\bar r)$ and  $f(\ell\bar r)$ is $f^2=F(2-k_0 \ell^{-2} F)$. Notice that $dF/d\bar r=f$ so that $F=\int_0^r{f(\bar r)d\bar r}$. To simplify the notation, we have set $\ell=1$ and removed the bar on $r$ in the main manuscript.

The metric \eqref{metricSS2b} resembles an FLRW metric, but it is important to examine this resemblance in a coordinate independent manner. Although the metric \eqref{metricSS2b} coincides with an FLRW metric at the center of symmetry $\bar r=0$ (so that  $L=1$ and $q^a=0$), the functional forms of $\rho$ and $p$ in \eqref{eqrhoCF1} and  \eqref{eqpCF1}  do not reduce to the FLRW equations for $\rho$ and $p$ (only the first two terms in \eqref{eqrhoCF1} and the first three terms in \eqref{eqpCF1} coincide with FLRW equations). 

Also, the geometric properties of the models with $k_0=0,-1$ in \eqref{metricSS2b} and \eqref{fFchi} might be very different from those of FLRW ``open'' models with $k_0=0,-1$. In these FLRW models  the proper length $\ell=a\int{dr}$ along radial rays (curves with fixed $t,\,\theta,\,\phi$ ) and the area distance ${\cal R}=\sqrt{g_{\theta\theta}}=a\,f$ both diverge as $r\to\infty$, but in the models \eqref{metricSS2b} and \eqref{fFchi} with $k_0=0,-1$ in \eqref{metricCF2}  both $\ell=a\int{dr/L}$ and ${\cal R}=a f/L$ converge in this limit (since $L\to\infty$), hence they are not ``open''.  Besides all these problems, the coordinate asymptotic limit $r\to\infty$ marks a curvature singularity (while $\rho,\,p\to 0$ in this limit, $q^a$ diverges in \eqref{eqQCF1}). Also,  as shown in \eqref{spatcurv}, the sign of the constant $k_0$ does not determine the spatial curvature in \eqref{metricSS2b}, which depends on $L$ and its gradients of $L$, which might diverge as $r\to\infty$. For all these reasons we did not consider the cases $k_0=0,-1$.

However, the ``closed'' models with $k_0=1$ share the main coordinate independent geometric features of closed FLRW models. The proper radial length $\ell$, area distance ${\cal R}$ are bounded and $L>0$ holds as long as $\epsilon_0$ is sufficiently small and $b=b(t)>0$ is bounded for all $t$. Also, these same conditions assure that spatial curvature in \eqref{spatcurv} is positive, since $f$ and $F$ in \eqref{fFchi} are bounded in the same coordinate range $0\leq r\leq \pi$ as in closed FLRW models. The models with $k_0=1$ admit 2 symmetry centers (at $r=0,\,\pi$) and the hypersurfaces of constant $t$ (rest frames) are conformal to $\mathbb{S}^3$. However, while the area distance ${\cal R}=af/L$ vanishes at the second symmetry center ($f(\pi)=0$), the metric function $L$ does not comply with the center regularity condition $L=1$ because $F(\pi)=2\ne 0$, but it complies with $L'=0$. 

\section{Accuracy of the series approximation}\label{sec:series}

\begin{figure}[h]
\centering
\includegraphics[width=0.8\linewidth]{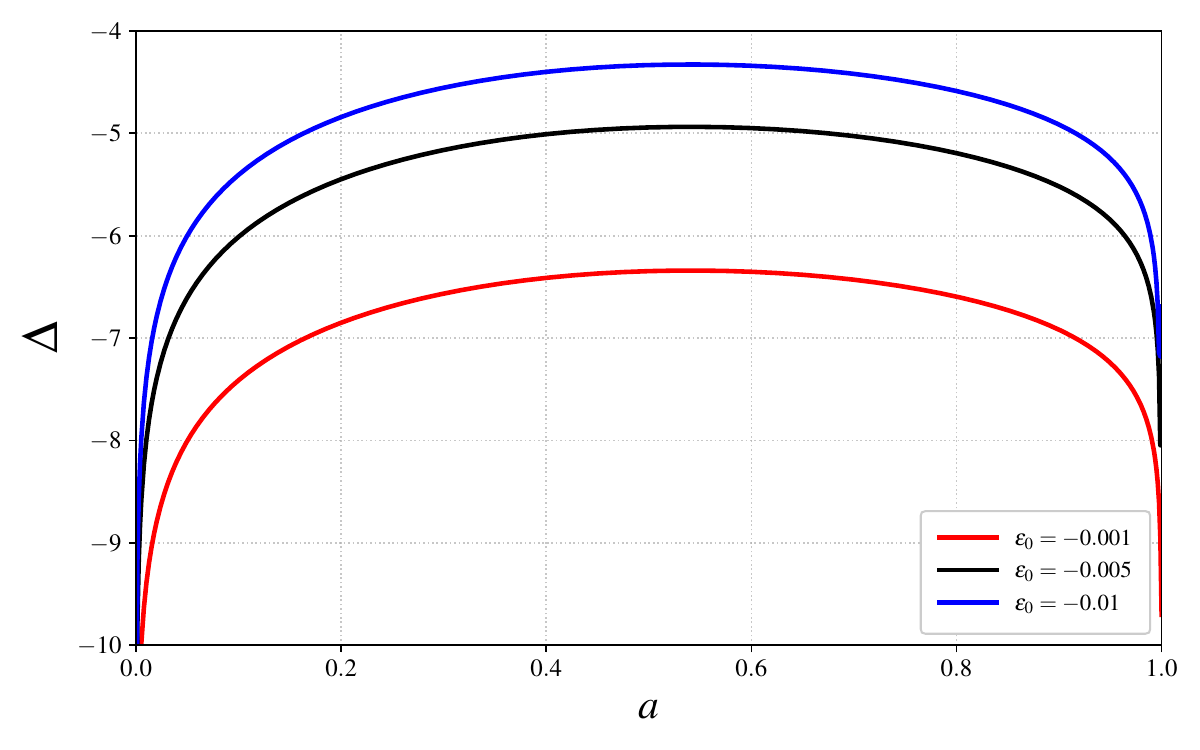}

\vspace{0.5cm}

\includegraphics[width=0.8\linewidth]{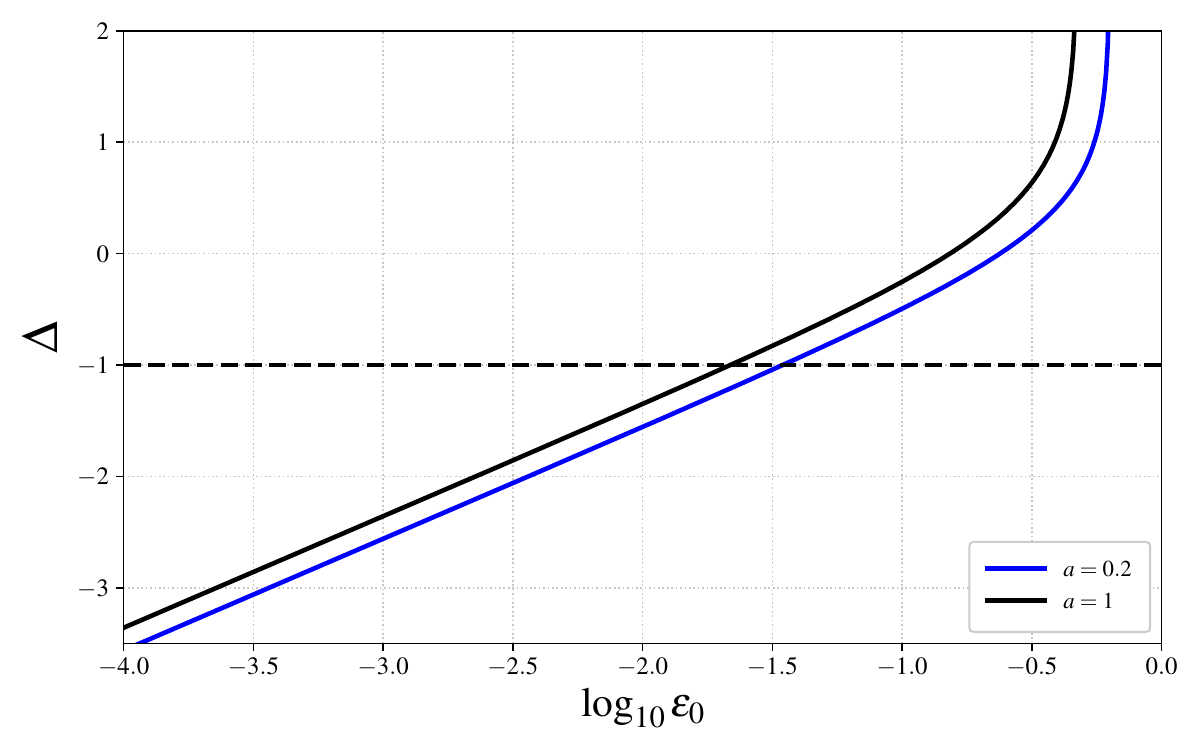}

\caption{%
Top: Ratio $\Delta=\log_{10}(\rho_{\tiny{\textrm{exact}}}/\rho_{\tiny{\textrm{approx}}}-1)$ as a function of $a$ for $r=\pi/2$ and various choices of $\epsilon_0$. 
Bottom: Ratio $\Delta$ as a function of $\epsilon_0$ and two choices of $a$. The values of $\epsilon_0$ below the dotted line ($\log_{10} \epsilon_0<-1.5$) denote the validity of the linear approximation of the series on $\epsilon_0$.}
\label{fig:figB}
\end{figure}

We show that the series expansion around $\epsilon_0$ in \eqref{eqrhoCF2}-\eqref{eqThCF2} provides a very accurate approximation to quantities computed with exact expressions \eqref{eqrhoCF2}-\eqref{eqQCF2}. We compare the total density $\frac{8\pi}{3H_0^2} \rho_{\tiny{\textrm{exact}}}$ from the exact equation \eqref{eqrhoCF1} and $\frac{8\pi}{3H_0^2} \rho_{\tiny{\textrm{approx}}}$ from the approximated equation \eqref{eqrhoCF2}. Top panel of Fig.~\ref{fig:figB} displays the ratio $\Delta=\log_{10}(\rho_{\tiny{\textrm{exact}}}/\rho_{\tiny{\textrm{approx}}}-1)$ as a function of $a$ for $r=\pi/2$, while the bottom panel of Fig.~\ref{fig:figB} displays $\Delta$ as a function of $\log_{10}\epsilon_0$ for $a=1$. In both graphs we used the parameters \eqref{Omegapars} and $\epsilon_0=0.005$. \\

\noindent

%Place Figures B1a and B1b\\

\noindent
We also compare the null geodesic curve $a=a(r)$ reaching the observer at $r=0,\,a=1$ (highlighted in top panel of Fig.~\ref{fig:fig5}) obtained as a numerical solution of the approximated differential equation \eqref{geod3} and the same curve from the solution of the exact differential equation
\begin{eqnarray}  
\left[\frac{da}{dr}\right]_{\textrm{null}}&=&\frac{
\mathcal{F}_1\,a\,\dot a
}{
(1+\epsilon_0e^{-a^2}F)\mathcal{F}_2},\nonumber\\
\label{nullgeod4}\\
\mathcal{F}_1&=&f\,F\, \epsilon_0^2\,e^{-2a^2}+(2(1+2a^2)F-\mu_0f)\epsilon_0 e^{-a^2}+2\,a\,\dot a\nonumber\\
\\
\mathcal{F}_2&=&-fF\epsilon_0^2 e^{-2a^2}+(2\mu_0 (1+2a^2)F\dot a-f)\epsilon_0e^{-a^2}+2\mu_0\dot a\nonumber\\
\end{eqnarray}
where $f,\,F,\,\dot a$ are given by \eqref{Fchi2} and \eqref{zeroO}. Bottom panel of Fig.~\ref{fig:figB} displays the ratio of $a(r)$ from the approximated solution to the exact solution. The difference between the curves is of the order of $10^{-3}$. 

 \end{appendix}
 
\section*{References}
\bibliographystyle{iopart-num}
\bibliography{referencesPV}

\end{document}